\documentclass{aa}  

\usepackage{graphicx}
\usepackage{xcolor}
\newcommand{\xmm}{{\it XMM-Newton}}

\newcommand{\nustar}{{\it NuSTAR}}
\newcommand{\chandra}{{\it Chandra}}

\def\gtsima{$\; \buildrel > \over \sim \;$}
\def\ltsima{$\; \buildrel < \over \sim \;$}
\def\gsim{\lower.5ex\hbox{\gtsima}}
\def\lsim{\lower.5ex\hbox{\ltsima}}

\usepackage{txfonts}
\begin{document}

   \title{Incoherent fast variability of X-ray obscurers}

   \subtitle{The case of NGC 3783}

   \author{B. De Marco
          \inst{1}\thanks{bdemarco@camk.edu.pl}
          \and
          T. P. Adhikari\inst{2,1}
          \and
          G. Ponti\inst{3}
          \and 
          S. Bianchi\inst{4}
          \and 
          G. A. Kriss\inst{5}
          \and 
          N. Arav\inst{6}
          \and 
          E. Behar\inst{7}
          \and 
          G. Branduardi-Raymont\inst{8}
          \and 
          M. Cappi\inst{9}
          \and 
          E. Costantini\inst{10}
          \and 
          D. Costanzo\inst{9}
          \and 
          L. di Gesu\inst{11}
          \and 
          J. Ebrero\inst{12}
          \and 
          J. S. Kaastra\inst{10,13}
           \and 
          S. Kaspi\inst{14}
           \and 
          J. Mao\inst{15,10}
          \and 
          A. Markowitz\inst{1,16}
           \and 
          G. Matt\inst{4}
          \and 
          M. Mehdipour\inst{10}
           \and 
          R. Middei\inst{4}
          \and 
          S. Paltani\inst{17}
          \and 
          P. O. Petrucci\inst{18}
          \and 
          C. Pinto\inst{19}
           \and 
          A. R\'o\.za\'nska\inst{1}
           \and 
          D. J. Walton\inst{20}
          }

   \institute{N. Copernicus Astronomical Center of the Polish Academy of Sciences,
              Bartycka 18, 00-716 Warsaw
         \and
         Inter-University Centre for Astronomy and Astrophysics (IUCAA), Ganeshkhind,  Pune - 411007, Maharashtra, India
         \and
         INAF-Osservatorio Astronomico di Brera, Via E. Bianchi 46, I-23807 Merate (LC), Italy
         \and
         Dipartimento di Matematica e Fisica, Universit\`{a} Roma Tre, Via della Vasca Navale 84, I-00146, Roma, Italy
         \and
         Space Telescope Science Institute, 3700 San Martin Drive, Baltimore, MD 21218, USA
         \and
         Department of Physics, Virginia Tech, Blacksburg, VA 24061, USA
         \and
         Department of Physics, Technion-Israel Institute of Technology, 32000 Haifa, Israel
         \and
         Mullard Space Science Laboratory, University College London, Holmbury St. Mary, Dorking, Surrey, RH5 6NT, UK
         \and
         INAF-IASF Bologna, Via Gobetti 101, I-40129 Bologna, Italy
         \and
         SRON Netherlands Institute for Space Research, Sorbonnelaan 2, 3584 CA Utrecht, The Netherlands
         \and
         Italian Space Agency (ASI), Via del Politecnico snc, 00133, Roma, Italy
         \and
         European Space Astronomy Centre, PO Box 78, 28691 Villanueva de la Caada, Madrid, Spain
         \and
         Leiden Observatory, Leiden University, PO Box 9513, 2300 RA Leiden, The Netherlands
         \and
         School of Physics and Astronomy and Wise Observatory, Tel Aviv University, Tel Aviv 69978, Israel
         \and
         Department of Physics, University of Strathclyde, Glasgow G4 0NG, UK
         \and
         University of California, San Diego, Center for Astrophysics and Space Sciences, 9500 Gilman Dr, La Jolla, CA 92093-0424, USA
         \and
         Department of Astronomy, University of Geneva, 16 Ch. d$'$Ecogia, 1290 Versoix, Switzerland
         \and
         Univ. Grenoble Alpes, CNRS, IPAG, 38000, Grenoble, France
         \and
         European Space Agency / ESTEC - Keplerlaan 1 - 2201 AZ, Noordwijk, Netherlands
         \and
         Institute of Astronomy, Madingley Road, CB3 0HA Cambridge, UK
             }

   \date{Received ...; accepted ...}

 
  \abstract
   {Obscuration events caused by outflowing clumps or streams of high column density and low ionised gas, shown to absorb the X-ray continuum heavily, have been witnessed across a number of Seyfert galaxies.}
   {We report on the X-ray spectral-timing analysis of the December 2016 obscuration event in NGC 3783, which was aimed at probing variability of the X-ray obscurer on the shortest possible timescales. The main goals of this study are to obtain independent constraints on the density and, ultimately on the distance of the obscuring gas, as well as to characterise the impact of variable obscuration on the observed X-ray spectral-timing characteristics of Seyfert galaxies.}
   {We carried out a comparative analysis of NGC 3783 during unobscured (using archival 2000-2001 \xmm\ data) and obscured states (using \xmm\ and \nustar\ data from the 2016 observational campaign). The analysed timescales  range between ten hours and about one hour. This study was then generalised to discuss the signatures of variable obscuration in the X-ray spectral-timing characteristics of Seyfert galaxies as a function of the physical properties of the obscuring gas.}
   {The X-ray obscurer in NGC 3783 is found to vary on timescales between about one hour to ten hours. This variability is incoherent with respect to the variations of the X-ray continuum. A fast response (on timescales shorter than about 1.5 ks) of the ionisation state of the obscuring gas to the short timescale variability of the primary X-ray continuum provides a satisfactory interpretation of all the observed X-ray spectral-timing properties. This study enabled us to put independent constraints on the density and location of the obscuring gas. We found the gas to have a density of $n_{e}> 7.1 \times 10^7 \rm{cm^{-3}}$, which is consistent with a location in the broad line region.}
{}


   \maketitle
%

\section{Introduction}

Flux variability is a common feature of active galactic nuclei (AGN; e.g. Padovani et al. 2017). 
It can be observed at any wavelength and over a very broad range of timescales (from minutes to years). Most of the variability observed at short wavelengths (from optical and UV to X-rays) is thought to originate within the accretion flow and to drive variability of reprocessed radiation (e.g. Guilbert \& Rees 1988; Blandford \& McKee 1982).
The time-dependent response of reprocessed radiation to the variable primary continuum can be used as a tool to map the environments of the black hole (BH). 
For example, optical/UV reverberation mapping techniques are routinely used to put constraints on the geometry and kinematics of the broad line regions (BLR; Blandford \& McKee 1982; Peterson 1993). \\
Thanks to the increasing availability of high signal-to-noise (S/N) data and long monitoring campaigns, a similar approach is now used to constrain the geometry of the innermost, X-ray-emitting regions of the accretion flow. These studies are based on the use of X-ray spectral-timing techniques (e.g. Vaughan \& Nowak 1997; Nowak et al. 1999; Wilkinson \& Uttley 2009; Zoghbi et al. 2010; Uttley et al. 2014). Such techniques are very powerful at singling out and constraining the causal relationship of spectral components contributing to X-ray variability on different timescales, that is, produced at different distances from the BH. Therefore, this analysis method allows us to map the close environments of BHs down to scales normally not accessible through standard techniques (e.g. De Marco et al. 2013, 2017; Kara et al. 2016, 2019).

So far, X-ray spectral-timing techniques have been mostly used to study the accretion flow around BHs. However, an additional important feature of BH activity is represented by outflows of photo-ionised gas, possibly associated with winds from the torus or the accretion disc (Koenigl \& Kartje 1994; Krolik \& Kriss 1995; Murray \& Chiang 1997; Proga et al. 2000; Blustin et al. 2005; Fukumura et al. 2010). These outflows manifest themselves as blue-shifted UV and X-ray absorption lines (e.g. Crenshaw et al. 2003). Depending on the ionisation state, column density and covering factor of the outflowing gas, these absorption features can greatly modify the observed emission from the AGN (e.g. Nardini et al. 2015; Reeves et al. 2018). 

AGN outflows might represent an important source of kinetic feedback. Constraining their physical properties, geometry, and production mechanism is crucial in determining their impact on the environment and the links between AGN activity and its host galaxy. AGN outflows are observed in at least half of type 1 nearby systems (e.g. Crenshaw et al. 2003; Cappi 2006; Tombesi et al. 2010;  Gofford et al. 2013; Laha et al. 2014; Parker et al. 2017a), from the more common warm absorbers, with speeds ranging from a few hundreds to a few thousands of km s$^{-1}$, to the ultra fast outflows, reaching speeds of $0.1-0.4c$. Although they are important for understanding AGN activity and feedback, these outflows are generally found to absorb only modest percentages of continuum flux as a consequence of their high ionisation level.

Recently, evidence has been reported for an additional class of outflows characterised by similar column densities 
($N_{\rm{H}}=10^{22-23}\rm{cm^{-2}}$) but typically lower ionisation parameter (${\rm{log}}\ \rm{(\xi/ erg\ cm\ s^{-1})} \lsim 2$). These so-called X-ray obscurers absorb significant amounts of UV and soft X-ray flux (Risaliti et al. 2011; Kaastra et al. 2014; Longinotti et al. 2013, 2019; Ebrero et al. 2016; Turner et al. 2018), partially eclipsing the X-ray source and producing broad, blue-shifted UV absorption troughs (Kaastra et al. 2014; Mehdipour et al. 2017, hereafter M17; Kriss et al. 2019, hereafter K19). Such obscuration events are transient (Markowitz et al. 2014), covering a diverse range of timescales, from a few hours (e.g. Risaliti et al. 2005, 2007, 2011), to days/months (Kaastra et al. 2018) or even years (Kaastra et al. 2014). 

A number of multiwavelength campaigns have allowed comprehensive spectral studies of obscuration events in AGN (e.g. Longinotti et al. 2013, 2019; Ebrero et al. 2016; Kaastra et al. 2014; M17). The main goal of these campaigns was to constrain the geometry and distance of the absorbing gas by studying the delayed response of the gas to variations of the ionising continuum.
However, standard time-resolved spectral analysis is affected by insufficient photon counts per time bin if the response timescale of the gas is very short. This problem can be overcome through the use of spectral-timing techniques, which measure the response of the gas by averaging over multiple variability cycles.
However, detailed spectral-timing studies of X-ray obscurers have not yet been attempted. The importance of the latter is two-fold. On the one hand, they can provide additional and independent constraints on the physical properties of the obscuring gas; in particular, by measuring the delay due to the response time of the absorbing gas to variations of the ionising continuum it is possible to estimate the electron density of the gas and, ultimately, its distance (e.g. Nicastro et al. 1999; Behar et al. 2003; Krongold et al. 2007; Kaastra et al. 2012; Silva et al. 2016). On the other hand, it is crucial to understand the impact of variable obscuration on the observed spectral-timing properties of the X-ray source. 
Indeed, the response timescales may be very short (Parker et al. 2017a; Pinto et al. 2018) and of the order of the timescales of intrinsic X-ray emission variability from the inner regions of the accretion flow. It is, therefore, of the utmost importance to correctly identify signatures of variable absorption in order to derive reliable constraints on the properties of the primary emitting regions. 
In particular, a basic prediction of radiative transfer theory is that the response of the absorbing gas to variations of the ionising flux is non-linear (e.g. Rybicki \& Lightman 1991). Therefore, variable absorption is expected to introduce an incoherent (i.e. non-linear; Vaughan \& Nowak 1997) fraction of variability in the energy bands most affected by absorption.

In this paper, we adopt an X-ray spectral-timing analysis approach to investigate the variability properties of the X-ray obscurer detected in the nearby ($z = 0.009730$, Theureau et al. 1998) Seyfert 1 galaxy NGC 3783 down to the shortest possible timescales. A multi-wavelength programme was triggered in December 2016 to study the source during the obscuration event which lasted for about a month (M17). Simultaneous optical-UV to X-ray observations allowed us to characterise the spectral properties of the obscurer (M17; Kaastra et al. 2018; Mao et al. 2019, hereafter Mao19; K19). This gas was found to have column density $N_{\rm{H}} \sim 10^{23}\rm{cm^{-2}}$, ionisation parameter ${\rm{log}}\ \rm{(\xi/ erg\ cm\ s^{-1})}=1.84^{+0.40}_{-0.20}$ , and outflow velocities up to $6000\ \rm{km\ s^{-1}}$, and to partially cover the X-ray source (with covering factor of $\sim 0.4-0.5$). In this paper, we make use of both archival and new X-ray data from the recent campaign (M17) to carry out a comparative analysis of the X-ray spectral-timing properties of NGC 3783 during unobscured and obscured epochs. 
We were able to study the fast variability of the X-ray obscurer down to timescales as short as about an hour and infer independent constraints on the density of the obscuring gas. A study of the short timescale variability properties of the reflected emission component will be presented in a companion paper (Costanzo et al. in prep.).

\section{Data reduction}
\label{sec:reduction}
We studied the X-ray spectral-timing properties of NGC 3783 using the two \xmm\ and joint \nustar\ observations of NGC 3783 carried out on 11 and 21 December 2016 as part of the campaign described in M17, and the three archival \xmm\ observations carried out on 28 December 2000, 17, and 19 December 2001. 
The 2000-2001 and the 2016 observations correspond, respectively, to unobscured (hereafter U1-U3) and obscured (herefter O1-O2) epochs of the source (M17, Kaastra et al. 2018). The log of these observations is reported in Table \ref{table1}.

\subsection{\xmm}
\label{sec:xmmred}
Given the higher effective area and the long uninterrupted exposures, for this analysis we used data from the EPIC pn detector only. Indeed, the EPIC MOS1 and MOS2 instruments experienced periods of Full Scientific Buffer throughout observations O1 and O2, thus precluding the use of these data sets. The EPIC pn was operated in Small Window mode during all the observations. The data reduction was carried out using the \xmm\ Science Analysis System (SAS v16.1), following standard procedures and with calibration files as of May 2018. The data were barycentre corrected.
For all \xmm\ observations, source counts were extracted from a circular region of $30$ arcsec. This extraction radius was chosen so as to reduce the impact of background flares affecting observations O1-O2\footnote{Given the PSF of the EPIC pn, the chosen extraction radius allows for a significant increase of the S/N in X-ray light curves of O1-O2 with respect to larger radii. For example, during O2 (the most affected by soft-proton flares) the net source count rate decreases by only 7\% with respect to an extraction radius of 45 arcsec, while the S/N increases by a factor of $\sim2$.} (see below). Background counts were extracted from two adjacent rectangular regions. We considered
only events with PATTERN $\leq$4.
Using the task \emph{epatplot} we verified that the source was not affected by pile-up during each observation.

Background-subtracted light curves were obtained using the SAS routine \emph{epiclccorr}. 
During observations U1-U3 soft-proton flares are either not present or observed at the end of each observation. These time intervals were removed, resulting in the effective exposures listed in Table \ref{table1}.
Observations O1 and O2 show soft-proton flares throughout the entire exposures. However, these contribute on average $\sim$13 percent of the measured count rate at the highest energies. Moreover, given their short duration (of the order of $\lsim 3$ ks), they are not expected to have significant contribution on the (longer) X-ray variability timescales tested. Therefore, we did not filter these events out, which resulted in two continuous observations of 110 ks and 55.5 ks exposures, respectively (see Table \ref{table1}). We verified that our results were not affected by this choice by comparing results obtained independently from the study of simultaneous \nustar\ data (see Sect. \ref{sec:fvar}).
Response matrices were obtained using SAS tasks \emph{rmfgen} and \emph{arfgen}. For consistency with Mao19 and M17, spectra of observations U1-U3 were combined using the task \emph{epicspeccombine}.

\subsection{\nustar}
The \nustar\ data were reduced using the standard pipeline included in the \nustar\ Data Analysis Software (NUSTARDAS) v1.8. and calibration files CALDB v20171204. Source counts were extracted from a circular region of 80 arcsec radius, and background counts from an adjacent circular region of 173 arcsec radius. The data were barycentre-corrected. The light curves from each of the two hard X-ray detectors FPMA and FPMB were extracted using the tool \emph{nuproducts}. The background-subtracted count rates from the two detectors were then summed within each time bin to increase the S/N.

\begin{table}
\centering
\caption{Log of the analysed observations. The table reports: (1) observation ID; (2)  observation date; (3)  nomenclature used throughout the paper to refer to the different datasets (U stands for `unobscured' , O stands for `obscured' ); (4) net exposure time (in the case of \xmm\ observations this is the exposure time after the removal of soft-proton flares).} 
\label{table1}
\begin{tabular}{lccc} 
\hline
 \noalign{\smallskip}
                   (1)    & (2) & (3)              &(4)                \\
                  ID & Start Time &      Obs        &  Exposure  \\
                        &    yyyy-mm-dd hh:mm:ss &      &    [ks]                \\
                 \noalign{\smallskip}
                \hline
                 \noalign{\smallskip}
                        \multicolumn{4}{c}{\xmm} \\
                 \noalign{\smallskip}
                \hline
                 \noalign{\smallskip}
                0112210101 & 2000-12-28 17:34:57 & U1 & 37 \\
                0112210201 & 2001-12-17 19:12:07 & U2  & 126 \\ 
                0112210501 & 2001-12-19 19:03:29 & U3 & 125 \\
                 \noalign{\smallskip}
                \hline
                 \noalign{\smallskip}
                0780860901 & 2016-12-11 09:15:48 & O1 &  110 \\ 
                0780861001 & 2016-12-21 08:36:16 & O2 &  55.5 \\
                 \noalign{\smallskip}
                \hline
                 \noalign{\smallskip}
                        \multicolumn{4}{c}{\nustar} \\
                         \noalign{\smallskip}
                \hline
                 \noalign{\smallskip}
                 80202006002 & 2016-12-11  21:56:08 & O1 & 55.6  \\
                 80202006004 & 2016-12-21 10:41:08 & O2 & 44  \\ 
                  \noalign{\smallskip}
                \hline
        \end{tabular}
\end{table}

\begin{figure*}
        \includegraphics[width=\textwidth]{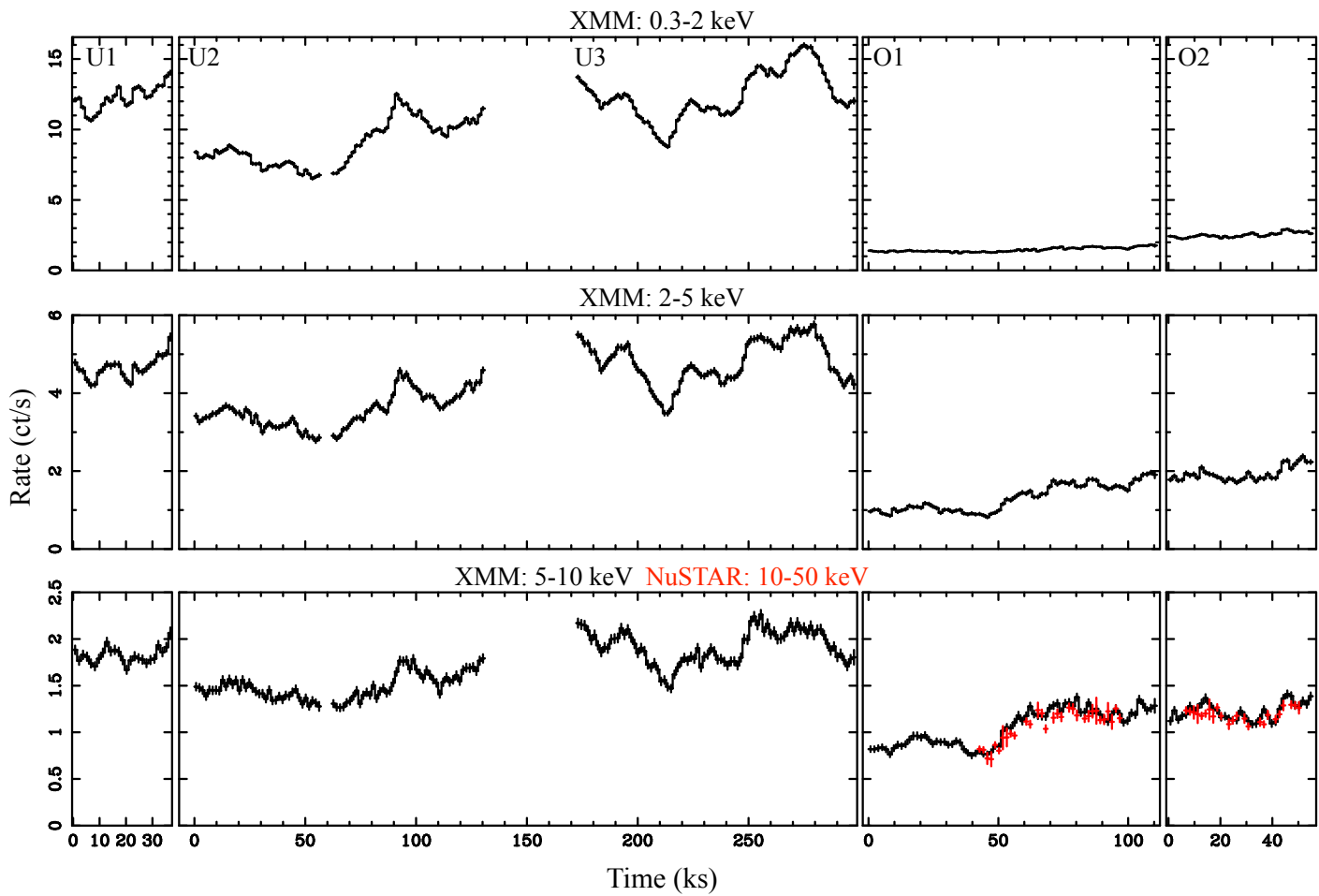}     
    \caption{Background-subtracted light curves of NGC 3783. The \xmm\ light curves are extracted in the $0.3-2$ keV, $2-5$ keV, and $5-10$ keV energy bands (black curves in upper, middle, and lower panels, respectively). The \nustar\ light curves are extracted in the $10-50$ keV energy band (red curves in the right-most lower panels). Note that in order to enable an easier comparison of the variability patterns in the \xmm\  $5-10$ keV and \nustar\ $10-50$ keV light curves, the \nustar\ light curves were not rescaled for the extraction area; thus, the match with \xmm\ is coincidental.}
    \label{fig:lc}
\end{figure*}

\begin{figure}
        \includegraphics[width=0.95 \columnwidth]{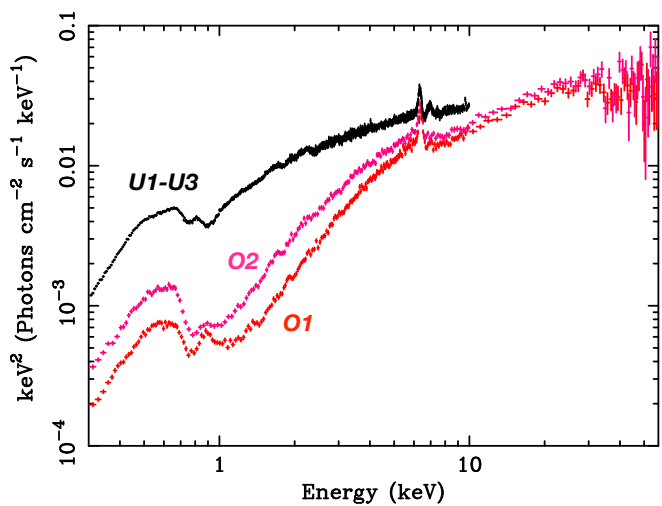}  
    \caption{X-ray spectra of NGC 3783 during unobscured (U1-U3) and obscured (O1-O2) epochs. Data for energies $E<10$ keV are from the \xmm\ EPIC pn detector, while data for energies $E>10$ keV during O1-O2 are from the \nustar\ FPMA detector.}
    \label{fig:spectrum}
\end{figure}

\section{Fast variability of the X-ray obscurer}
\label{sec:var}

\subsection{Light curves}
Fig. \ref{fig:lc} shows the background-subtracted \xmm\ and \nustar\ light curves of NGC 3783 in the energy bands: $0.3-2$ keV, $2-5$ keV, $5-10$ keV, and $10-50$ keV. The light curves have a time bin of 1.5 ks. 
The source shows remarkable variability on the sampled timescales. A severe drop in the $0.3-2$ keV count rate (a factor of $\sim 5-6$) characterises observations O1-O2 with respect to the archival observations U1-U3. This is due to the mildly ionised (${\rm{log}}\ \rm{(\xi/ erg\ cm\ s^{-1})}= 1.84$), high column density ($N_{\rm{H}}\sim10^{23} \rm{cm^{-2}}$) obscuring outflow, detected during the latest observations (M17; Kaastra et al. 2018; K19). Fig. \ref{fig:spectrum} shows the effects of the obscuration event on the X-ray spectra of the source. The intensity of the drop decreases at higher energies (a factor of $\sim$3 in the $2-5$ keV band and a factor of $\sim$1.6 in the $5-10$ keV band) as the fraction of transmitted X-ray flux through the absorbing gas increases (see figure 5 of M17). The \nustar\ $10-50$ keV (red points in Fig. \ref{fig:lc}, last two panels) and \xmm\ $5-10$ keV light curves follow the same trend of variability during the joint observation period, suggesting a lack of major spectral variations associated with the primary hard X-ray continuum.\\

\subsection{$\rm{F}_{var}$ spectrum}
\label{sec:fvar}
We first investigated the short timescale spectral variability of the source in a model independent way by computing the fractional root mean square (\emph{rms}) variability amplitude spectrum ($\rm{F}_{var}$; e.g. Vaughan et al. 2003; Ponti et al. 2004, 2006). When referring to a short timescale, we mean timescales of the order of a few hours. We used light curves split into segments of 36 ks length to this aim. The light curves were extracted in adjacent energy bins, with a time bin of 200 s, and background-subtracted.

For the \xmm\ datasets, we computed the Poisson noise-subtracted periodogram (normalised to units of squared fractional \emph{rms}, Miyamoto et al. 1991) of each segment. We 
integrated it over the frequency range $\sim 2.78\times 10^{-5} - 3.3\times 10^{-4}$ Hz, which corresponds to timescales of ten to about one hour, and from its square root we obtained an estimate of the $\rm{F}_{var}$ from each segment. $\rm{F}_{var}$ estimates associated with each observing epoch (U1-U3 and O1-O2) were then averaged together. 

In the case of observations O1-O2 we extended the analysis of the $\rm{F}_{var}$ spectrum up to $\sim$ 50 keV using the available \nustar\ data. In order to sample the same realizations of the underlying variability process with both instruments, we selected only time intervals of strict simultaneity between the \xmm\ and \nustar\ exposures (see Fig. \ref{fig:lc}, bottom panel). 
Given the presence of gaps due to Earth occultation during \nustar\ observations, the $\rm{F}_{var}$ of \nustar\ data was computed by measuring the light curves variance in the time domain. In order to sample the same range of timescales as for the $\rm{F}_{var}$ spectrum of \xmm\ data, we used segments of 36 ks and a time bin of 1.5 ks. The energy bins were chosen so as to ensure a minimum of 25 counts within each time bin.
Results are shown in Fig. \ref{fig:fvar}. It is worth noting that in their common energy band, the \xmm\ and \nustar\ $\rm{F}_{var}$ spectra are in good agreement (black and grey squares in the right panel of Fig. \ref{fig:fvar}), confirming that the background flares (see Sect. \ref{sec:xmmred}) in the \xmm\ data do not significantly affect the analysed timescales.

The $\rm{F}_{var}$ spectrum of the source shows significant spectral variability within each epoch as well as between the two epochs. Observations U1-U3 are characterised by a mildly decreasing trend of variability amplitude with energy. A narrow feature is observed between $\sim 6-7$ keV, corresponding to the energy of the neutral Fe K$\alpha$ line. This feature hints at the presence of a reflection component which does not vary on timescales shorter than ten hours, as expected if produced by distant material (Sect. \ref{discussion:ngc3783}).

During observations O1-O2, the source displays a drop of fractional variability (a factor of $\sim$4) in the soft X-ray band ($E\lsim$ 1.5 keV). This is due to the presence of constant (on the sampled timescales) scattered emission from distant gas (as revealed in the time-averaged spectra of the source, Mao19). This component dominates in the soft energy band because of the significant absorption blocking the emission from the central regions. At $E\gsim$3 keV the $\rm{F}_{var}$ shows the same decreasing trend of variability amplitude as a function of energy as observed during U1-U3. 
The feature ascribable to a constant narrow Fe K$\alpha$ line is still visible.

\subsection{Modelling of the $\rm{F}_{var}$ spectrum}
\label{sec:modfvar}

\begin{figure*}
        \includegraphics[angle=0,width=\textwidth]{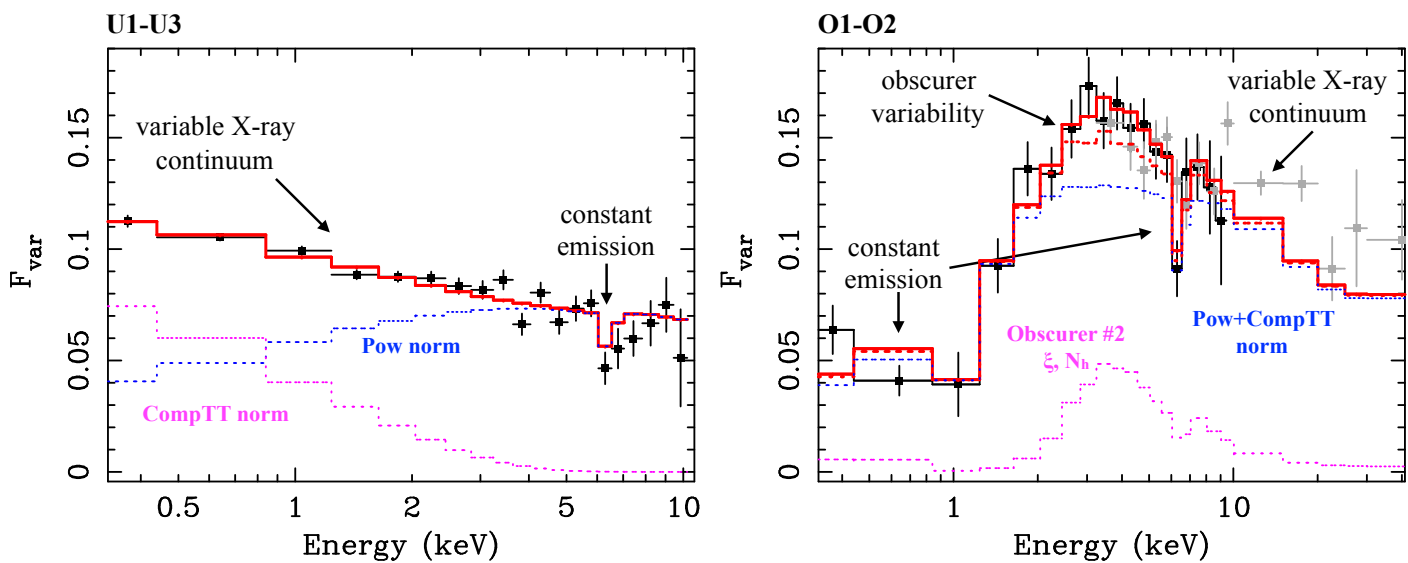}
        
    \caption{$\rm{F}_{var}$ spectra of observations U1-U3 (left panel) and O1-O2 (right panel). Black and grey squares refer, respectively, to \xmm\ and \nustar\ data. Blue and magenta dashed curves show $\rm{F}_{var}$ models obtained by letting different parameters (as indicated in the labels) of the continuum or of obscurer \#2 vary, while red solid curves are obtained by letting all those parameters vary simultaneously. Red dashed curve in the right panel shows the simple model obtained assuming only variations of the primary continuum and correlated variations of the ionisation parameter of obscurer \#2. The main spectral features contributing to the shape of the $\rm{F}_{var}$ spectra at different energies are schematically indicated by arrows in the plots.}
    \label{fig:fvar}
\end{figure*}

Results presented in M17 and Mao19 from the analysis of the X-ray spectra of NGC 3783 during U1-U3 and O1-O2 revealed high spectral complexity. As seen in the $\rm{F}_{var}$ spectra, this complexity is coupled with significant spectral variability (Fig. \ref{fig:fvar}).
We investigated the contribution of the different spectral components identified in those works to the observed short timescale variability. 
To this aim we created $\rm{F_{var}}$ models, and compared them to the observed $\rm{F_{var}}$ spectra reported in Sect. \ref{sec:fvar}.\\

$\rm{F_{var}}$ models were obtained by making use of the best-fit models to the mean energy spectra presented in M17 and Mao19. All the absorption and scattered emission components were modeled using table models produced with the spectral synthesis code Cloudy (v17.01, Ferland et al. 2017) and the SEDs calculated in M17. Fits were performed within Xspec v12.10\footnote{We note that M17 and Mao19 performed spectral modelling using \texttt{SPEX} (Kaastra et al. 1996). The use of Cloudy tables within Xspec might yield systematic differences of the order of 20\%. However, these are not significant for our scopes, given the low spectral resolution of the data analysed in this paper.} (Arnaud 1996).
 
The models include the following spectral components common to both the unobscured and obscured datasets (with the corresponding Xspec model name within brackets): Galactic absorption (\texttt{phabs}); Comptonised hard X-ray band emission (\texttt{cutoffpl}); Comptonised soft X-ray band emission (\texttt{compTT}); two scattered emission components from distant gas (Cloudy table models); X-ray reflection from distant gas (\texttt{pexmon}); three warm absorber components, and one high ionisation absorption component (Cloudy table models).

To account for the obscurer during O1-O2, we added two partially covering, mildly ionised absorption components (\texttt{partcov} convolved with Cloudy table models) with ionisation parameter fixed at the value ${\rm{log}}\  \rm{(\xi/ erg\ cm\ s^{-1})}= 1.84$ obtained from combined X-ray and UV diagnostics (K19) and covering fraction as determined in M17.

We note that Mao19 reports the detection of nine phases of the warm absorber and three scattered emission components from the fit of high resolution RGS and \chandra\ HETGS spectra.
However, CCD-resolution spectra (and the even lower resolution of $\rm{F_{var}}$ spectra) do not allow us to resolve all the different phases as most of them are characterised by small differences in outflow velocity and ionisation parameter. Therefore, we used only three absorption components to model the warm absorber (forcing their ionisation parameter to vary within the ranges determined in Mao19, i.e. ${\rm{log}}\ \rm{(\xi/ erg\ cm\ s^{-1})}= -0.6$ to 3.1) and two scattered emission components. Nonetheless, the inferred total $N_{\rm{H}}$ is broadly consistent with that obtained by Mao19.

We reran the fits to allow the model parameters to adjust. For consistency with previous papers (M17; Mao19) and with the aim of applying the same models, observations O1 and O2 were fit simultaneously while observations U1-U3 were combined into a single spectrum. The parameters left free to vary are: the spectral slope of the power-law, the normalization of all emission components, the column density of the two mildly ionised obscurers, the ionisation parameter and column density of the high-ionisation absorption component, and the parameters of the three warm absorber components. All the other parameters were fixed at the best-fit value obtained in M17 and Mao19. The parameter values inferred from these fits are reported in Appendix \ref{sec:app} and Table \ref{table2}.

In order to reproduce the observed $\rm{F_{var}}$ spectra, relevant parameters of the best-fit spectral model were allowed to vary randomly following a normal distribution centred on their best-fit value. The width of the distribution determines the variability amplitude of each spectral parameter. We tested different width values so as to recover the observed $\rm{F_{var}}$ spectral shape. For the normalization of continuum emission components (\texttt{cutoffpl} and \texttt{compTT}) we assumed a log-normal distribution, since this kind of distribution is found to well describe the X-ray flux distribution of AGN  (Uttley et al. 2005; also see Alston et al. 2018 and Alston 2019). 
The models that qualitatively describe the observed $\rm{F_{var}}$ spectra better are shown as solid and dotted lines in Fig. \ref{fig:fvar}.\\

Over the course of observations U1-U3, the shape of the $\rm{F_{var}}$ spectrum can be explained in terms of short timescale variability (between about one hour and ten hours)  of both the soft and hard X-ray Comptonisation components (Fig. \ref{fig:fvar}, left). 
Indeed, variations of the normalization of the \texttt{cutoffpl} component\footnote{The cut-off power-law component is normalised to the flux in the $7-10$ keV energy range.} (blue curve in Fig. \ref{fig:fvar}, left)  closely resemble the observed $\rm{F_{var}}$ spectrum at $E\gsim$3 keV, but leave excess variability in the soft X-ray band if the soft X-ray Comptonisation component is assumed to be constant. 
Concurrent variations of the normalization of the soft X-ray Comptonisation component (magenta curve in Fig. \ref{fig:fvar}, left) can account for the soft X-ray band shape of the $\rm{F_{var}}$ spectrum during observations U1-U3. Agreement between the $\rm{F_{var}}$ model and the data is found assuming variations of $\sim$16\%--19\% and 6\%--7\%, respectively, for the normalization of the soft and hard Comptonisation components.

During observations O1-O2, variability of the soft and hard X-ray Comptonisation components (assuming variations of their normalizations of $\sim$ 10\%--11\% and 13\%--15\%, respectively; blue curve in Fig. \ref{fig:fvar}, right) can account for most, but not all, of the observed variability. This model describes the hard X-ray $\rm{F_{var}}$ spectrum well (from $E\sim$ 7 keV) up to the energies covered by \nustar\ (Fig. \ref{fig:fvar}, right). The drop of variability power in the soft X-ray band is produced by the constant (on the sampled timescales) scattered emission component, dominating this part of the spectrum, as the obscurer blocks the emission from the innermost regions.
However, the peak of variability observed at $\sim3$ keV is not recovered by a model which includes only variable continuum emission as the latter underpredicts the amount of fractional variability at those energies. Nonetheless, this peak can be recovered by allowing for short timescale variability of the X-ray obscurer.
Fig. \ref{fig:fvar} (dashed red curve) reports a simple model where variations of the ionisation parameter, $\xi$, of obscurer \#2 are related to variations of the power law flux, $F_{pow}$, by $\xi \propto F_{pow}$, with the scaling factor derived from the time-averaged best-fit model. The model reproduces the observed spectral shape well. Additional variations (by $\sim 13-15$\%) of the column density of obscurer \#2 cannot be excluded and would further contribute to the observed variability peak (Fig. \ref{fig:fvar}, continuous red curve and dashed magenta curve). In Appendix \ref{sec:app2}, we show that the observed $\rm{F_{var}}$ spectrum can be also reproduced assuming variations of $N_{\rm{H}}$ alone, possibly combined with either mild variations of covering factor of obscurer \#2 (Fig. \ref{fig:fvar_nhcf}), or variations of the parameters of obscurer \#1. However, a model describing the observed variability in terms of variations of the ionisation parameter of obscurer \#2 related to variations of the ionising continuum, is consistent with results presented in the following sections.

Our analysis shows that the obscuring gas in NGC 3783 varies on timescales in the range between one hour and ten hours. From a physical point of view, the observed variations could be due to a fast response of the obscuring gas to variations of the illuminating continuum, possibly combined with variability associated with an inhomogeneous structure crossing our line of sight.
In any event, changes in the properties of the obscurer are expected to produce non-linearly correlated (i.e. incoherent) variability (e.g. Rybicki \& Lightman 1991). To verify this hypothesis, we carried out a comparative analysis of the \emph{rms} and covariance spectra of the source, and of the coherence spectra.

\section{Incoherent variability of the X-ray obscurer}
\label{sec:incoh_var}

\subsection{Rms and covariance spectra}
\label{sec:freqresolved}
Spectral fits of the \emph{rms} spectrum allow for an identification of all components variable on the sampled range of timescales (e.g. Revnivtsev et al. 1999; Gilfanov et al. 2000), while the covariance spectrum pinpoints which of those components are linearly correlated (or coherently variable, Wilkinson \& Uttley 2009; Uttley et al. 2014). 
Thus, a discrepancy between the \emph{rms} and the covariance spectrum indicates that a non linearly-correlated component is contributing in that specific energy range.
Contrary to the $\rm{F_{var}}$ spectrum, these spectra are not modified by constant (on the sampled timescales) spectral components, thus reducing the number of parameters to be included in the fits.

\begin{figure*}
        \includegraphics[angle=0,width=\columnwidth]{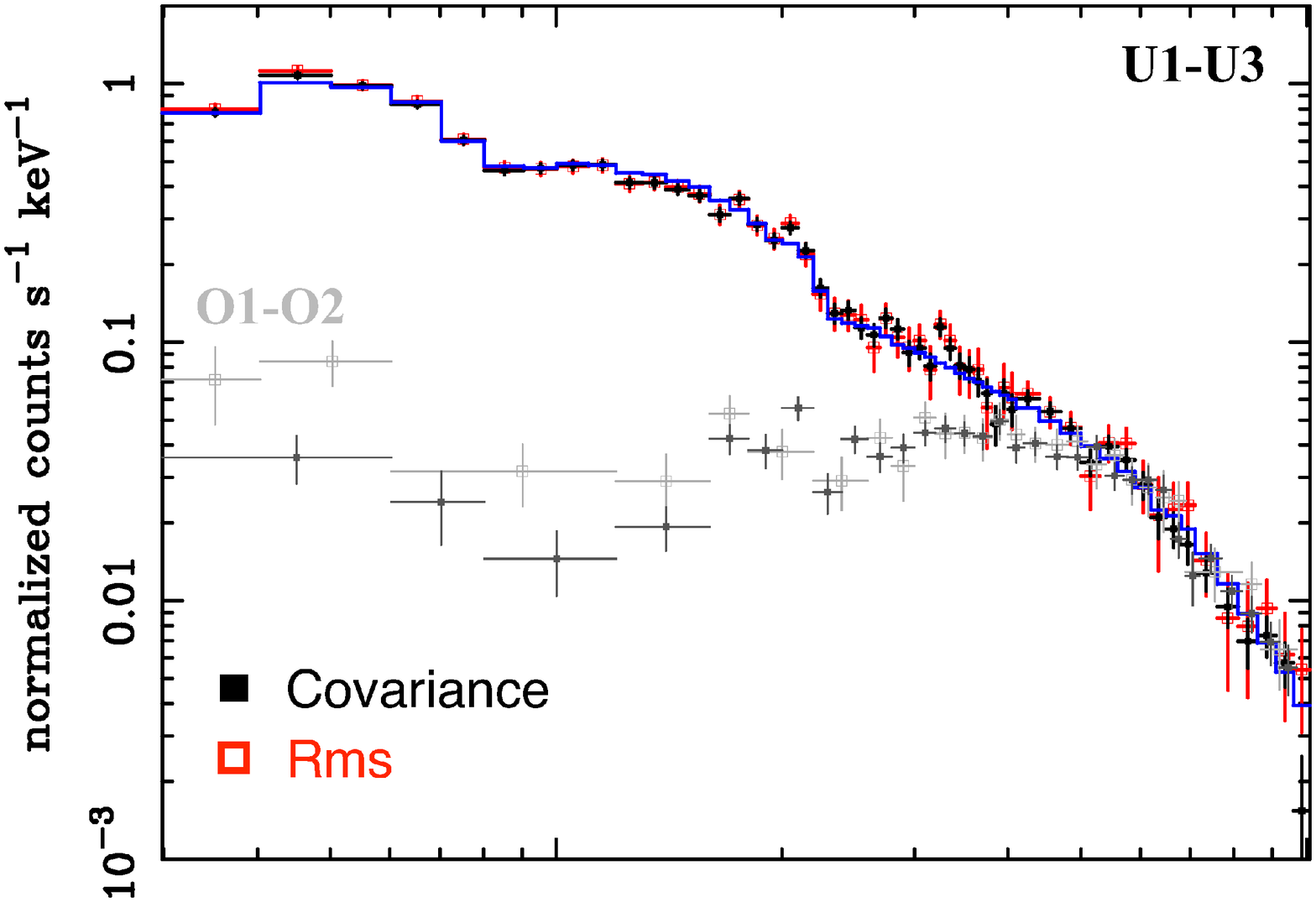}  
        \hspace{0.6cm}  
        \includegraphics[angle=0,width=\columnwidth]{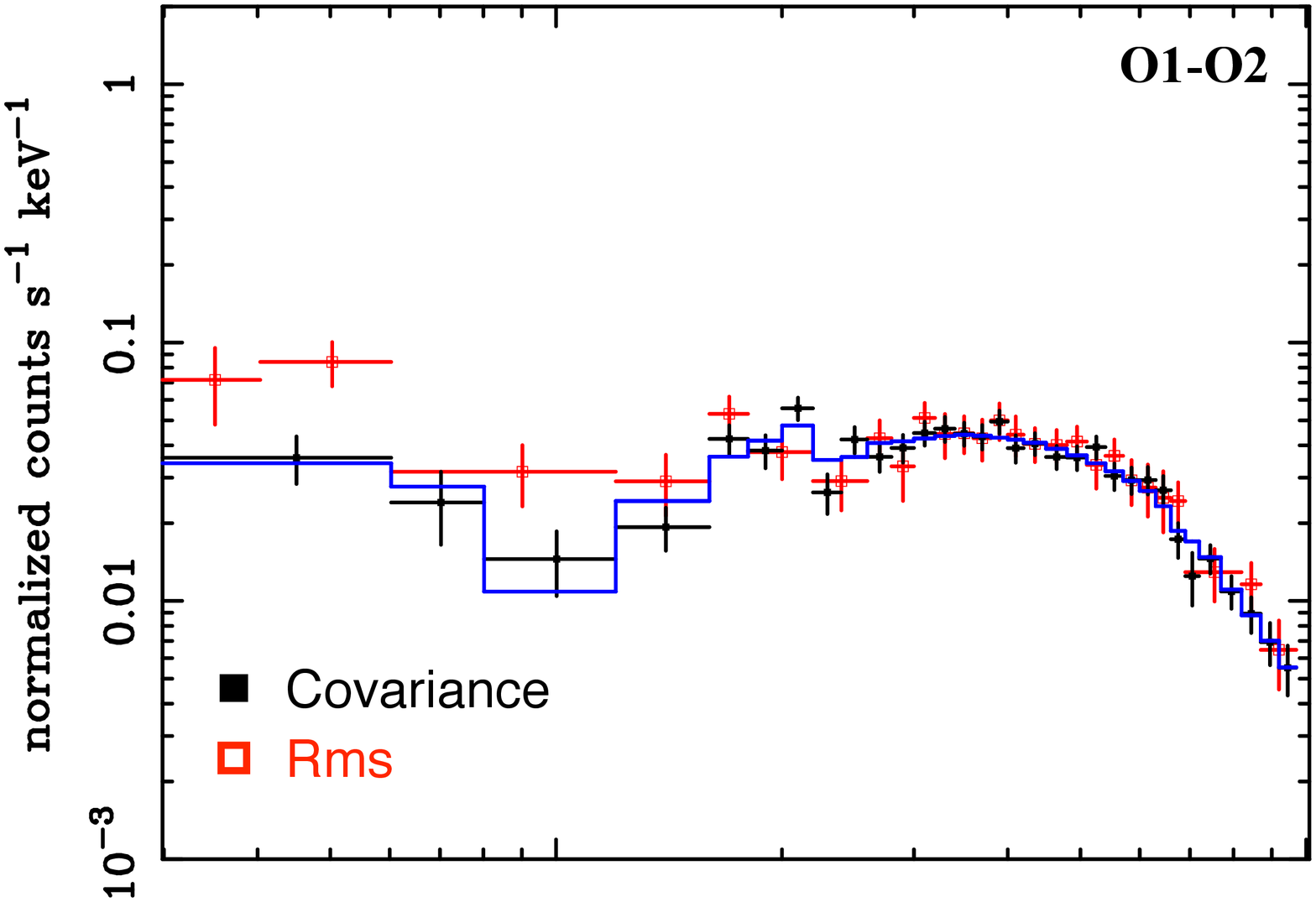}  
        \vspace{0.1cm}
        \hspace{0.06cm} 
        \includegraphics[angle=0,width=\columnwidth]{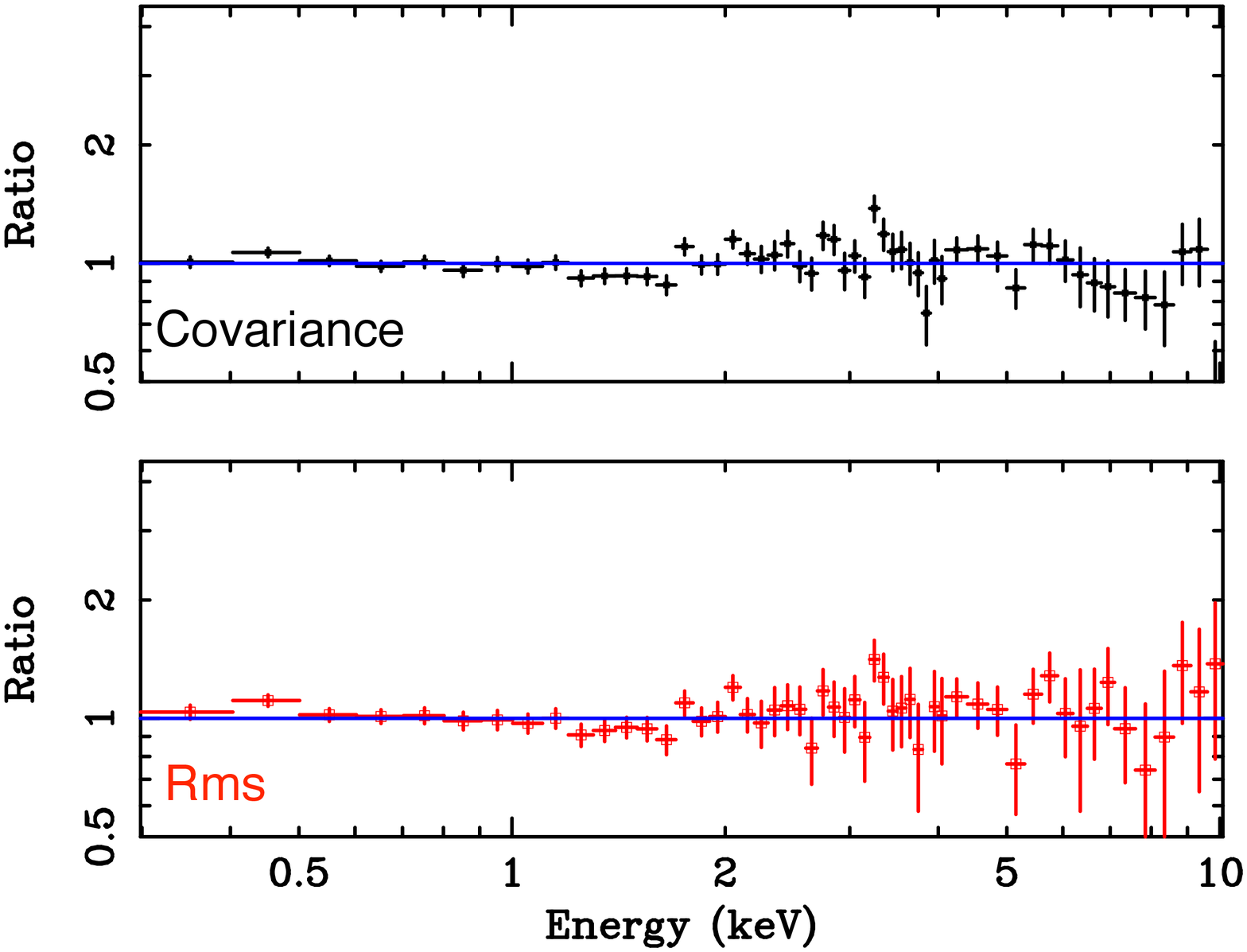}  
        \hspace{0.6cm}  
        \includegraphics[angle=0,width=\columnwidth]{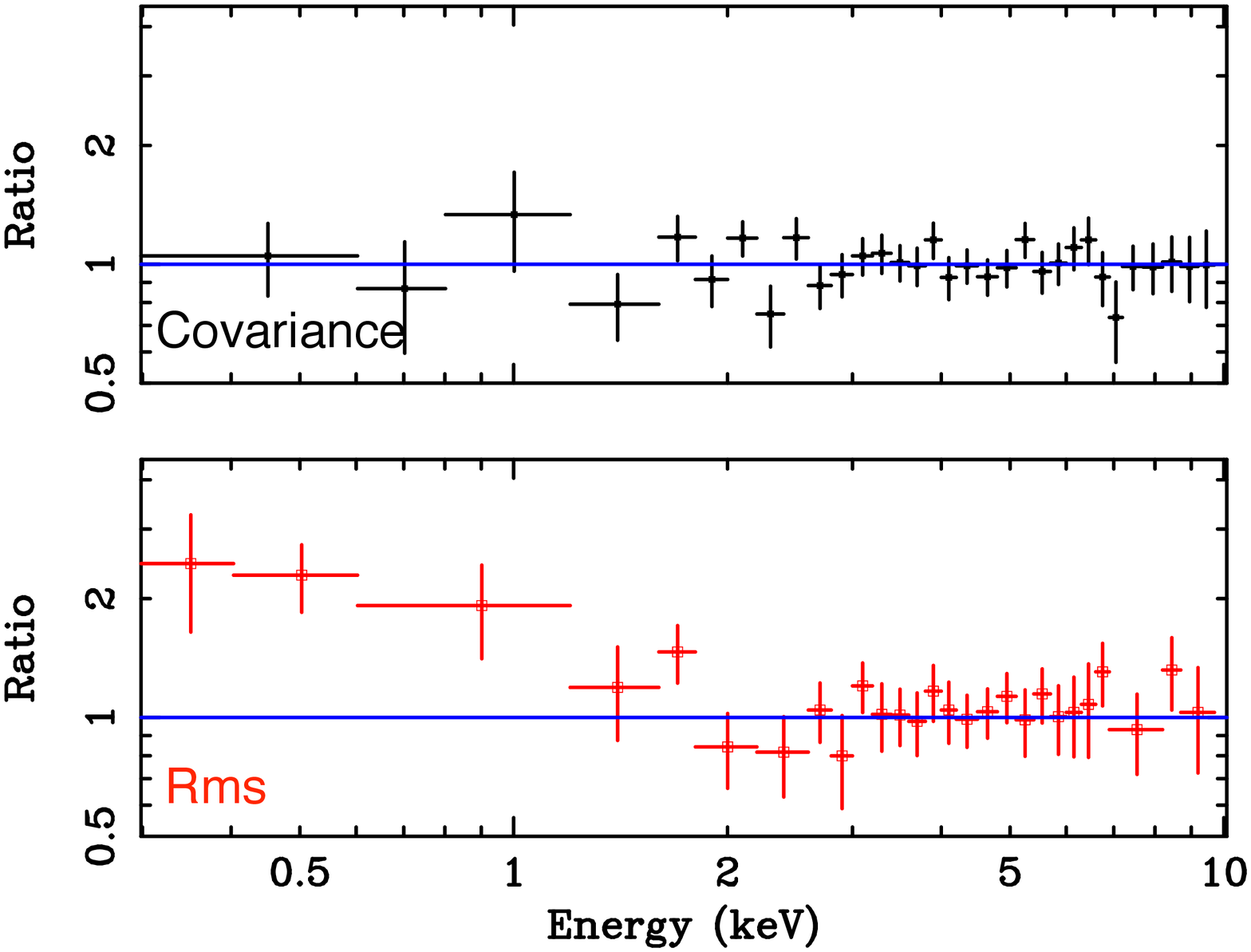}  

    \caption{\emph{Rms} and covariance spectra of U1-U3 (left panels) and O1-O2 (right panels). Blue curve is the best-fit model to the covariance spectrum. \emph{Rms} and covariance spectra of O1-O2 are also overplotted on those of observations U1-U3 (left panel, grey squares) for an easier comparison. The bottom panels show the ratio of \emph{rms} and covariance spectra to the best-fit model of the covariance spectrum.}
    \label{fig:covarrms}
\end{figure*}

The \emph{rms} spectrum is computed following the same procedure used for the $\rm{F_{var}}$ spectrum (Sect. \ref{sec:fvar}) but adopting absolute counts units normalization for the periodogram (e.g. Vaughan et al. 2003). The covariance spectrum is computed as \begin{math} Cov\ (\nu,E_i)=\sqrt{\Delta\nu\ (\mid \bar{C}_{R,E_i}(\nu)\mid^2-n^2)}/\bar{P}_{R}(\nu)\end{math}, where $\bar{C}_{R,E_i}(\nu)$ is the mean Fourier cross-spectrum between a reference band and adjacent energy bins (the contribution of each energy bin is subtracted from the reference band in order to remove correlated Poisson noise due to photons contributing in both bands; see Uttley et al. 2014) over the frequency interval $\Delta\nu$, $n^2$ is a bias term due to the contribution of Poisson noise to the modulus-squared of the cross spectrum (see Uttley et al. 2014), and $\bar{P}_{R}(\nu)$ is the mean Poisson-noise subtracted periodogram of the reference band. We used the 0.3-10 keV energy band as the reference band. 
To be consistent with the $\rm{F_{var}}$ spectrum, also for the analysis of \emph{rms} and covariance spectra we focused on the frequency range $\Delta
\nu \sim 2.78\times 10^{-5} - 3.3\times 10^{-4}$ Hz, corresponding to timescales of ten hours down to one hour.
The \emph{rms} and covariance spectra of the source are shown in Fig. \ref{fig:covarrms} for the two epochs.

The \emph{rms} and covariance spectra are perfectly matched during U1-U3 (Fig. \ref{fig:covarrms}, left), implying that all variable spectral components are also linearly correlated with the reference band. In this same figure we overplot the \emph{rms} and covariance spectra (in light and dark grey, respectively) of observations O1-O2, to highlight the differences between the two epochs. Notably, all the spectra overlap at high energies (at $\gsim$ 5 keV), meaning that the spectral-timing properties of the primary hard X-ray continuum did not change between the two sets of \xmm\ observations (15 years apart). Major differences are instead seen below $\sim$ 5 keV, due to the presence of the X-ray obscurer during observation O1-O2 (see below). In addition, during O1-O2 (see Fig. \ref{fig:covarrms}, right panel), the \emph{rms} spectrum shows an excess in the soft X-ray band with respect to the covariance spectrum. This indicates the presence of additional variability not linearly correlated with the broad band X-ray continuum. 

The blue continuous curves in the left and right panel of Fig. \ref{fig:covarrms} show the best-fit models ($\chi^2/d.o.f.=$114/73) obtained from the simultaneous fit of the covariance spectra during the two epochs. For these fits we used a model comprising all the components listed in Sect. \ref{sec:modfvar} (see also Appendix \ref{sec:app}) for the corresponding epoch, apart from the scattered emission and the distant reflection components. Indeed, those components are found to be constant on the timescales tested here (see Sect. \ref{sec:fvar}), thus they would not imprint any signature in the \emph{rms} and covariance spectra. On the other hand, we included all the absorption components reported in Sect. \ref{sec:modfvar}, because emission components are modified by absorption in both the \emph{rms} and covariance spectra irrespective of whether the absorption component is variable or constant (e.g. Ar\'{e}valo et al. 2008; Bhayani \& Nandra 2010). We left the normalization and spectral index of the primary power-law, and the normalization of the soft Comptonisation component free to vary. The other parameters were fixed at the best-fit values of the time-averaged spectrum (Table \ref{table2}), apart from the covering factor and the column density of the two obscurers. These parameters were allowed to vary between the best-fit values obtained from the simultaneous fit of O1 and O2 (see Table \ref{table2}). The quality of the fit does not improve if the constraints on the covering factor and column density of the two obscurers are removed.

In the bottom panels of Fig. \ref{fig:covarrms}, for each epoch we show the ratios of the covariance and the \emph{rms} spectra to the best-fit model obtained from the simultaneous fit of the covariance spectra. Our simple model including a (absorbed) cutoff power-law plus a soft Comptonisation component aptly describes both the \emph{rms} and covariance spectra during U1-U3. This means that the observed short term variability can be entirely ascribed to these two components, with their variations being linearly correlated. On the other hand, the best-fit model to the covariance spectrum of observations O1-O2, while still requiring the presence of variable and linearly correlated power-law and soft Comptonisation components, leaves some excess residual in the \emph{rms} spectrum at $E\lsim 1.5$ keV. Assuming that the properties of the soft Comptonisation component did not change between the two epochs (i.e. it still varies coherently with the X-ray continuum during O1-O2; thus, it does not contribute to the soft band excess variability in the \emph{rms} spectrum), then these residuals can only be accounted for by letting the parameters of the obscurers free to vary. Though the improvement in the fit is not significant because of the low S/N of the spectra ($\Delta\chi^2\sim11$ for $\Delta d.o.f.=6$), this gives indications, in agreement with theoretical expectations, that the obscurer is the cause of the observed incoherent excess variability. This inference is supported by the lack of incoherent variability during unobscured epochs. In Sect. \ref{sec:coherE}, we further test this hypothesis.

\subsection{Coherence}
\label{sec:coherE}
Our analysis of $\rm{F_{var}}$, \emph{rms}, and covariance spectra (Sects. \ref{sec:fvar}-\ref{sec:modfvar} and \ref{sec:freqresolved}) show the presence of incoherent short timescale variability during observations O1-O2 (Fig. \ref{fig:fvar} and Fig. \ref{fig:covarrms}) likely associated with the obscurer. 
We investigated through simulations whether changes of the ionisation state of the obscurer can result in the production of incoherent variability in the soft band on timescales of a few hours.

To this aim we first measured the intrinsic coherence $\gamma^{2}_{I}$ (Vaughan \& Nowak 1997; Uttley et al. 2014) of the source as a function of energy, in the two epochs.
While the $\rm{F_{var}}$ (Sect. \ref{sec:fvar}) shows the distribution of variable flux over energy, the coherence spectrum picks out only the coherently variable (i.e. linearly correlated with the broad band X-ray continuum) fraction, and is not influenced by the presence of constant components.

We followed the procedure extensively described in Uttley et al. (2014), whereby the intrinsic coherence is computed as \begin{math} \gamma^{2}_{I} (\nu,E_i) = [\mid \bar{C}_{R,E_i}(\nu)\mid^2-n^2]/[\bar{P}_{R}\bar{P}_{E_i}(\nu)]\end{math}, with all the terms having the same definition as described in Sect. \ref{sec:freqresolved}. Since this relies on the use of Fourier-domain techniques, we did not extend the analysis to the \nustar\ band as previously done for the $\rm{F_{var}}$ spectrum, because the gaps present in \nustar\ data would bias such analysis. 
As for the covariance spectrum, we chose the $0.3-10$ keV band as the reference band, and the same frequency range for the computation of average cross and power spectra ($\sim 2.78\times 10^{-5} - 3.3\times 10^{-4}$ Hz; i.e. we tested coherent variability on timescales of ten hours down to one hour).  
For the computation of the errors on $\gamma^{2}_{I}$ we used equation 8 of Vaughan \& Nowak (1997), which accounts for both the uncertainty in the Poisson noise and in the intrinsic coherence.

\begin{figure*}
        \includegraphics[angle=0,width=\columnwidth]{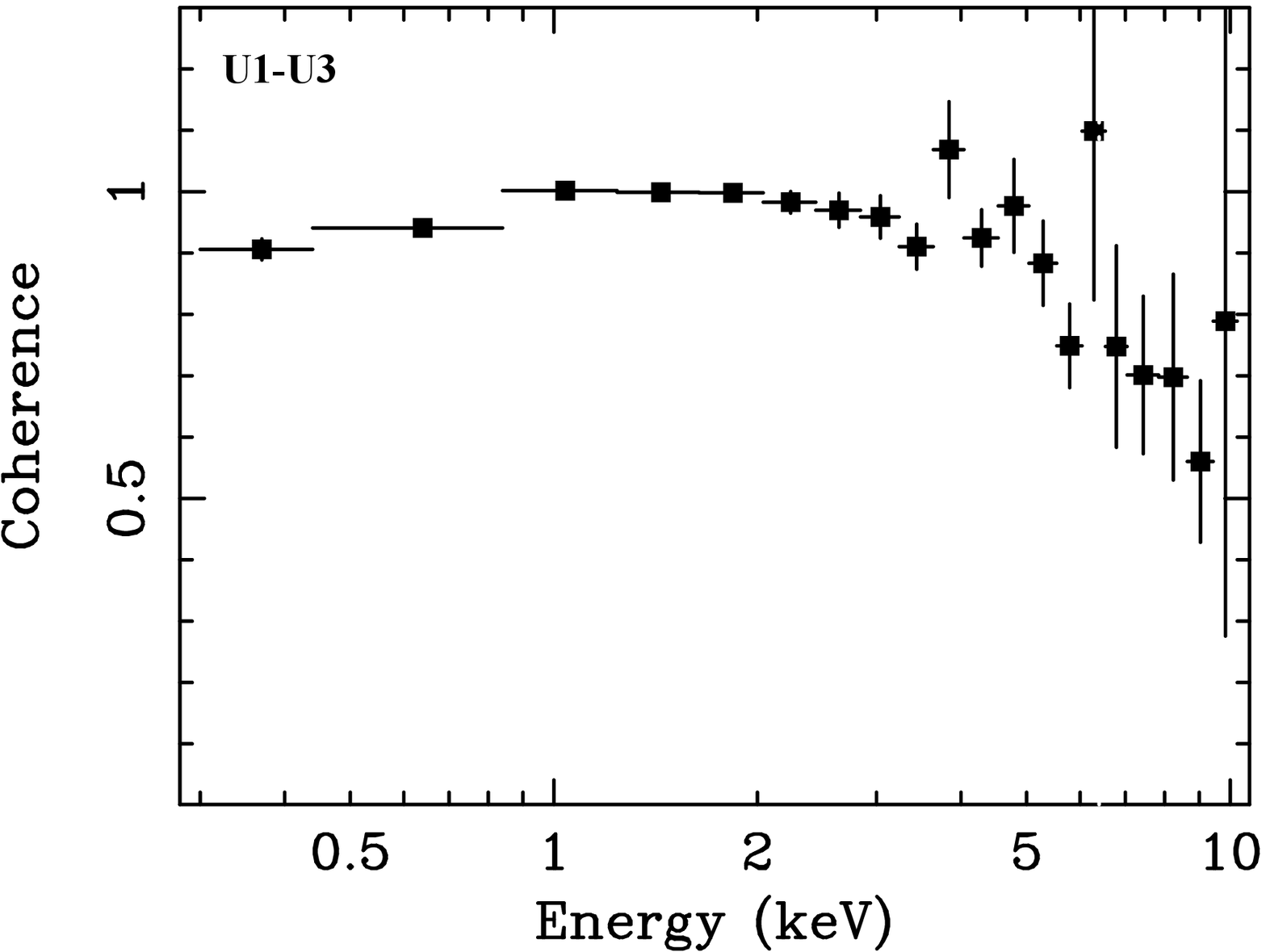} 
        \hspace{0.6cm}
        \includegraphics[angle=0,width=\columnwidth]{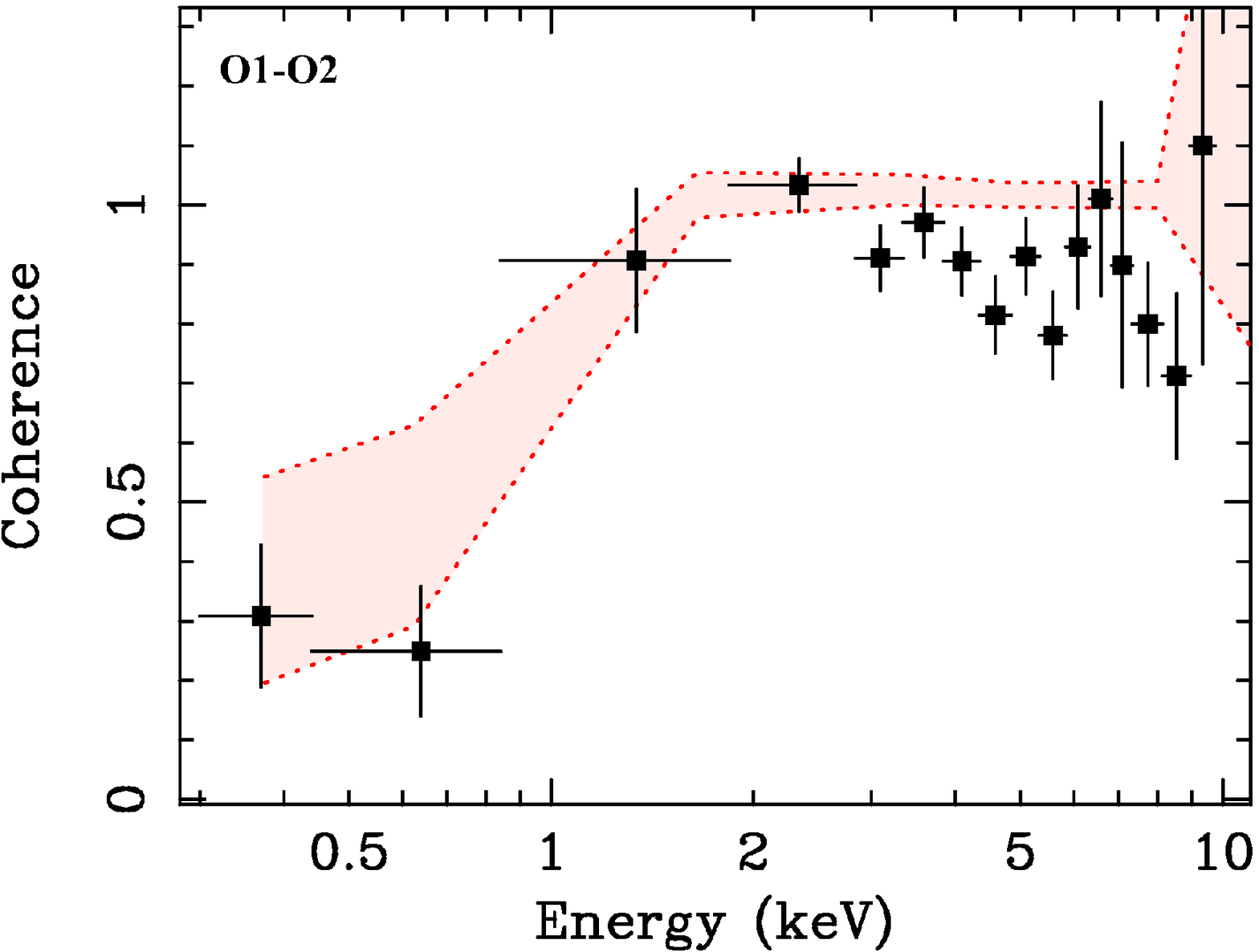}
    \caption{Coherence as a function of energy (U1-U3 left panel; O1-O2 right panel) over the frequency range $2.78\times10^{-5}-3.3\times10^{-4}$ Hz, and with the $0.3-10$ keV band as the reference band. Red shaded area in the right panel marks the 90\% confidence level contours obtained from Cloudy simulations of an obscuring gas located at a distance of 10 light days, of electron density $n_{e}=2.6\times10^{9}\rm{cm^{-3}}$ , and log $N_{\rm{H}} = 23.1$, responding to the observed variations of the ionising continuum on timescales $\tau_{rec}< 1500$ s. The ionisation parameter of the simulated obscuring gas varies within the range ${\rm{log}}\ \xi =1.53-1.82$ as a consequence of short timescale variations of the ionising continuum.}
    \label{fig:coherence}
\end{figure*}

The intrinsic coherence is shown in Fig. \ref{fig:coherence}. As expected from results presented in Sect. \ref{sec:freqresolved}, we detected high coherence at all energies during observations U1-U3.
The coherence is maximum at intermediate energies (between $\sim$ 1 and 5 keV) and moderately decreases towards lower and higher energies. 
This behaviour is seen in several AGN (Epitropakis \& Papadakis 2017, see also Sect. \ref{discussion:continuum}), whereby the intrinsic coherence decreases with the energy separation between bands.\\

Observations O1-O2 show similar levels of coherence at $E\gsim 2$ keV, again suggesting that the intrinsic spectral-timing properties of the primary X-ray continuum did not undergo significant changes between the two epochs. 
However, the coherence drops (by a factor of $\sim3$) below $\sim 1$ keV, due to uncorrelated variability components significantly contributing in this band, as manifested by the excess soft band residuals seen in the \emph{rms} spectrum (Sect. \ref{sec:freqresolved}). However, the fact that the intrinsic coherence is not consistent with zero in the soft X-ray band means that residual contribution from linearly correlated variability is present, as indeed inferred from the covariance spectrum (Sect. \ref{sec:freqresolved}).

Therefore, we carried out simulations to test whether variations of the ionisation state of the obscurer can explain the observed drop of coherence (and, consequently, also the excess soft band residuals in the \emph{rms} spectrum) during observations O1-O2. 
To this aim, we used Cloudy to simulate the spectra of an obscured, variable power-law, following the observed variations of the primary X-ray continuum during observations O1-O2. 
We created 111 synthetic spectra of the transmitted plus scattered emission produced by an obscuring gas responding to variations of the primary continuum by changing its ionisation state. Each spectrum corresponds to an integration time of 1.5 ks, covering the entire duration of observations O1-O2. We assumed that the gas reaches photo-ionisation equilibrium with the ionising continuum within each time bin of 1.5 ks. This assumption is justified by the detection of short timescale variability (i.e. on timescales longer than about one hour) of the obscurer (Sect. \ref{sec:fvar}), meaning that the gas must respond on very short timescales, and is in agreement with the tentative indication of a soft lag possibly associated with the response of the gas as will be discussed in Sect. \ref{sec:lagfreq}.
Using the parameters derived in M17, we simulated an obscuring gas with a column density of log $N_{\rm{H}} = 23.1$ (corresponding to the average value inferred by M17), electron density of $n_e=2.6\times 10^9 \rm{cm^{-3}}$ (this value is consistent with the assumed response time of $<$1.5 ks, see discussion in Sect. \ref{discussion:ngc3783}), and solar abundances. The corresponding depth of the absorbing material is $\Delta d=4.8\times10^{13}$ cm. Given the assumed density, the measured average ionisation parameter and the luminosity of the source (i.e. $L_{1-1000\ Ryd}\sim 0.6-1.1\times10^{44}\ \rm{erg\ s^{-1}}$, see Sect. \ref{discussion:ngc3783}), the gas should be located at a maximum distance of $\sim$ 10 light days. Here we focus on results obtained for a distance of 10 light days of the absorbing gas from the ionising source, since this better reproduces the observed coherence, but we refer to Sect. \ref{discussion:obsc_inSey1} for a more general discussion. The simulations make use of the unobscured (i.e. corrected for obscuration) SED derived in M17 from the 2016 data. For each simulated spectrum, the entire SED was rescaled so that the $5-10$ keV luminosity follows the variations observed in this energy band, and assuming the continuum varies only in normalization (as also inferred from our modelisation of the $\rm{F_{var}}$ spectra, Sect. \ref{sec:modfvar}). The choice of referring to the $5-10$ keV energy band to determine the range of variations of the ionisation parameter is dictated by the need to obtain light curves which are free from the contribution of the obscurer short-term variability (Fig. \ref{fig:fvar}). Of course, the entire 1-1000 Ryd contributes to the variability of the ionisation parameter, but the omitted low energy part is expected to either increase or not change (depending on whether or not it varies on short time scales) the range of variations of the ionisation parameter assumed for our simulations. We verified that by artificially increasing the range of variations of the ionisation parameter the results presented here do not change significantly. Finally, we also tested the more physical assumptions of a SED variable on the short time scales tested only at energies $E>13.6$ eV, or only at $E>0.1$ keV, finding consistent results in all the cases. Intrinsic luminosities were obtained assuming a distance of 38.5 Mpc for the source (Tully \& Fisher 1988).

We obtained energy-dependent light curves of the transmitted plus scattered emission by integrating each synthetic spectrum within the energy bins used to compute the coherence spectrum (Fig. \ref{fig:coherence}, right).
Fluxes were then converted to observed count rate, and randomised in order to include the effects of counting noise.
We used these light curves to compute the expected coherence spectrum resulting from variability of the ionisation state of the obscuring gas in response to variations of the ionising continuum. This is shown in Fig. \ref{fig:coherence} (right), where the corresponding 90 percent confidence contours (red shaded area) are overplotted on the data. 

Our simulations show that changes in the ionisation state of the obscurer can account for the production of incoherent variability, reproducing the observed drop of coherence at soft energies. For a distance of 10 light days of the absorbing gas from the X-ray source, given the observed variability of the source and the corresponding range of spanned ionising luminosities, the ionisation parameter of the simulated gas varies between ${\rm{log}}\ \xi =1.53-1.82$ (consistent with the range of values $1.84^{+0.40}_{-0.20}$ estimated by K19). These simulations recover well the amplitude of the observed drop of coherence.

\begin{figure}
        \includegraphics[angle=0,width=0.95\columnwidth]{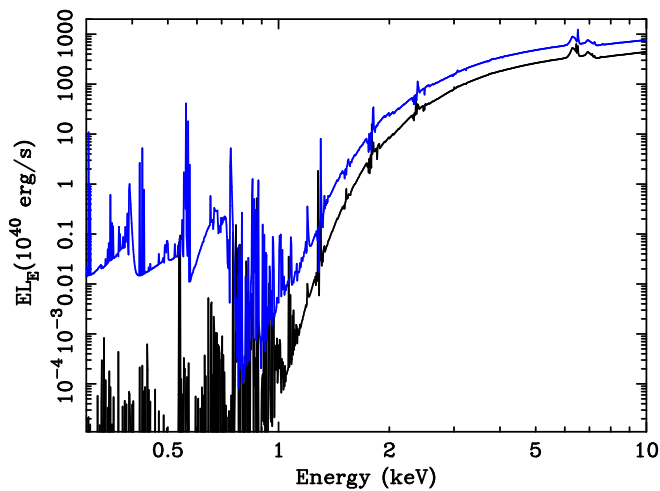} 
    \caption{Examples of two Cloudy synthetic spectra showing the transmitted plus scattered emission from an obscuring gas, responding to the observed variations of X-ray continuum. The two spectra are from the two time bins during observations O1 (black) and O2 (blue), respectively corresponding to the lowest and highest input ionising luminosity. These spectra do not include the contribution of the constant scattered emission component produced by distant material.} 
    \label{fig:sims}
\end{figure}

Fig. \ref{fig:sims} shows two of the simulated Cloudy spectra. Note that the simulated spectra do not include the constant scattered emission component from distant gas, which dominates at low energies, and causes the decrease of $\rm{F_{var}}$ seen in the data (Fig. \ref{fig:fvar} right panel). The model predicts complex variability in the soft band associated with the response of the obscurer. Due to photo-ionisation and recombination processes, the response to continuum variability across the soft energy band is not linear. This is a consequence of fundamental radiative transfer theory, in that the effects of an absorbing gas on the incident radiation is described by an exponential function (e.g. Rybicki \& Lightman 1991), so that variations of the ionisation parameter and/or column density of the gas lead to non-linear variations of the absorbed flux.
Such non-linear modulation of continuum variability by the absorbing gas introduces an incoherent variability component. This is responsible for the significant loss of coherence observed in the soft band (Fig. \ref{fig:coherence}, right panel). 
Nonetheless, the net transmitted flux in the soft band follows the variations of the primary continuum. Therefore, not all the coherence is destroyed. In particular, since the transmitted fraction increases if the gas is more ionised and/or if the gas layer is thinner, the fraction of coherent-to-incoherent variability (and thus, the intensity of the drop) depends on the ionisation state of the gas and on its depth (see Sect. \ref{discussion:obsc_inSey1}).  
As a matter of fact, the measured coherence does not drop to zero in the data.

Finally, it is important to note that the model considered here does not include the effects of partial-covering gas. According to partial covering models, a fraction of the primary continuum does not intercept the obscurer, thus its contribution has to be added to the transmitted flux of the remaining fraction which is intercepted by the obscurer. Nonetheless, the fraction of primary continuum that does not intercept the obscurer is intrinsically highly coherent, therefore, its contribution significantly increases the coherence in the soft band. We verified that allowing for even very small fractions of unobscured continuum would substantially reduce the drop of coherence seen in the soft band, preventing us from reproducing the data. For a partial covering model to reproduce the coherence spectrum of the source a more complex modelisation including an additional source of incoherent variability (e.g. from the second obscurer) would be needed. This aspect is further discussed in Sect. \ref{discussion:ngc3783}\\ 
In conclusion, our results support changes of the ionisation state of the obscurer as the cause of the detected short timescale incoherent variability.

\subsection{Lag-frequency spectrum}
\label{sec:lagfreq}

\begin{figure*}
        \includegraphics[angle=0,width=0.95\columnwidth]{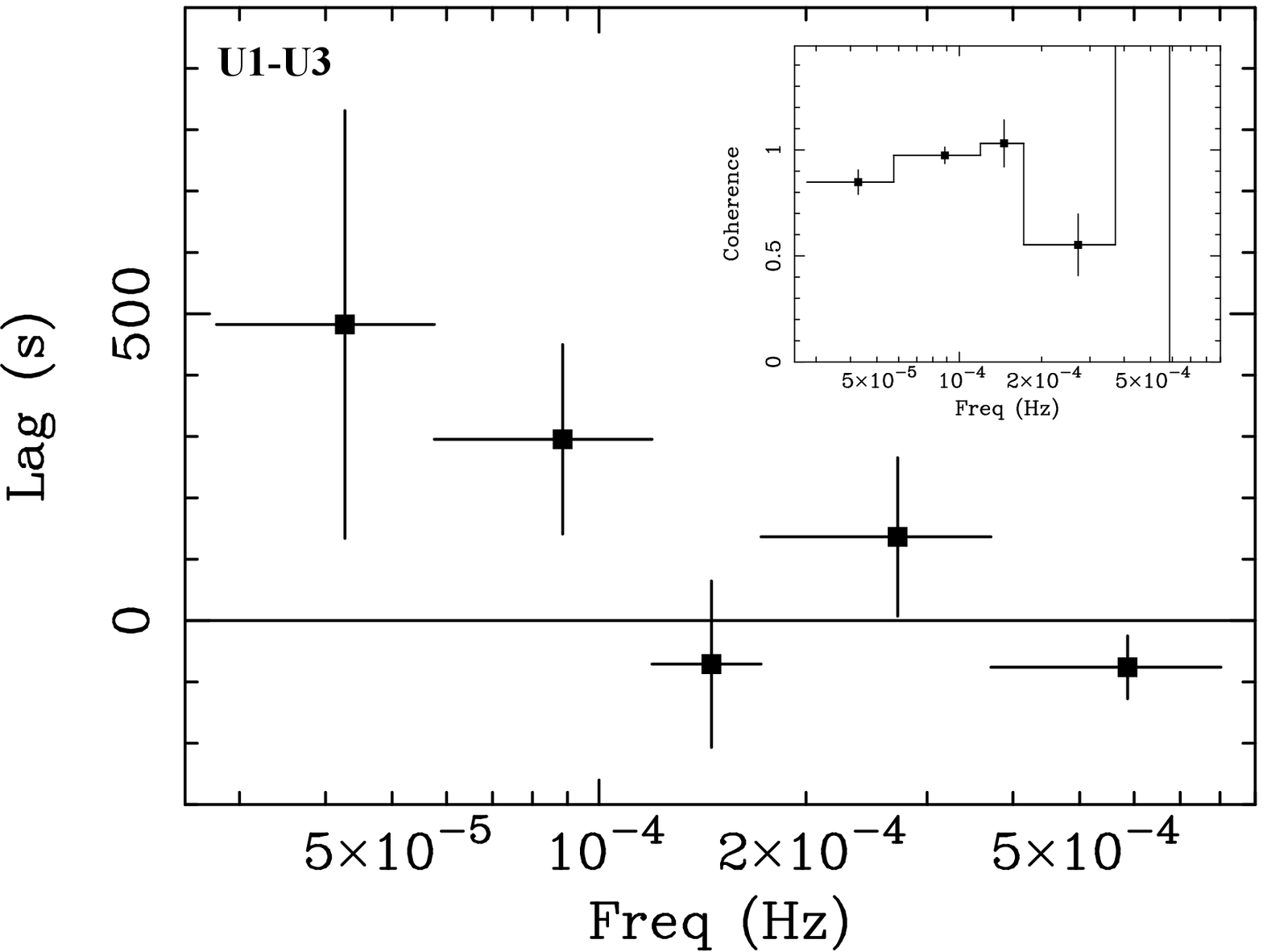} 
        \hspace{0.6cm}
        \includegraphics[angle=0,width=0.95\columnwidth]{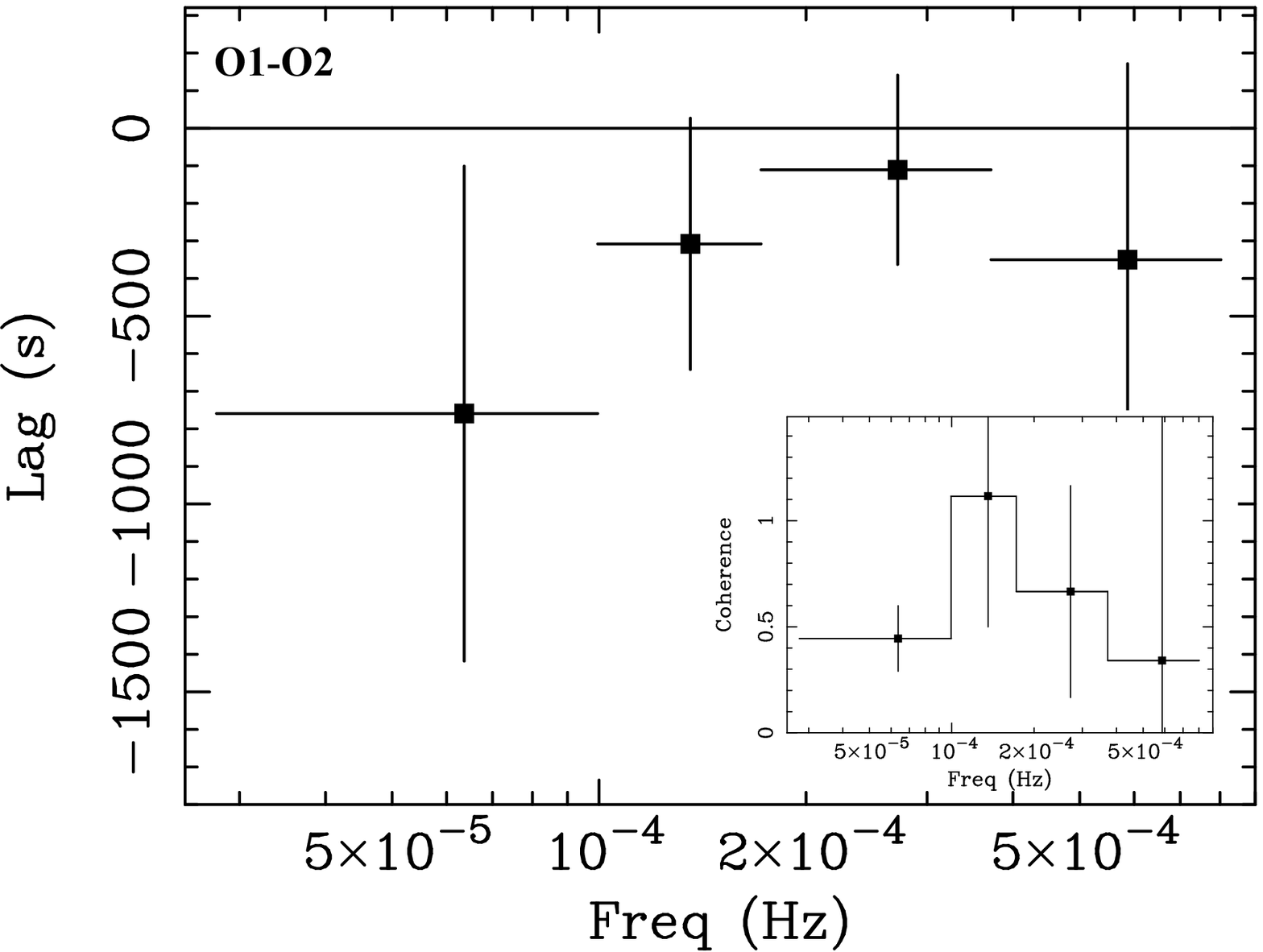}
    \caption{Lag-frequency spectrum of observations U1-U3 (unobscured, left panel) and of O1-O2 (obscured, right panel) between the $0.5-2$ keV and $4-10$ keV energy bands. The insets show the intrinsic coherence as a function of frequency between the two energy bands.}
    \label{fig:lagfreq}
\end{figure*}

We looked for the presence of time delays in the data associated with a response of the ionisation state of the gas to variations of the ionising continuum.
To this aim, we analysed time delays between the soft ($0.5-2$ keV) and the hard X-ray band ($4-10$ keV). The former is the energy band dominated by absorption during O1-O2, while the latter is dominated by the primary X-ray continuum (see Fig. 2 in M17). The range 2-4 keV is excluded from the analysis in order to avoid ambiguity due to the presence of significant contribution from both components.
Time lags were computed as $\tau(\nu) = \phi(\nu)/2\pi\nu$, where $\phi(\nu)$ is the frequency-dependent phase of the average cross-spectrum between soft and hard band light curve segments (Uttley et al. 2014), and rebinned using a multiplicative rebinning factor of 1.2. Results are shown in Fig. \ref{fig:lagfreq} for both unobscured and obscured epochs so as to allow for a comparison. A positive (negative) lag indicates a delayed response of hard (soft) photons with respect to soft (hard) photons. 

During observations U1-U3, a hard lag is observed at low frequencies ($\nu\sim10^{-4}$ Hz). This kind of lag is commonly observed in AGN (De Marco et al. 2013; Walton et al. 2013; Lobban et al. 2014; Kara et al. 2016; Epitropakis \& Papadakis 2017; Papadakis et al. 2019) and can be ascribed to delays intrinsic to the X-ray continuum (see discussion in Sect. \ref{discussion:continuum}). In this dataset we do not observe signatures of soft X-ray lags. Soft X-ray lags can be ascribed to reflected or thermally reprocessed radiation from the innermost accretion flow (so-called X-ray reverberation, Fabian et al. 2009; De Marco et al. 2013), or to a delayed response of an absorber component (Silva et al. 2016). 
However, given the black hole mass of the source ($M_{\rm{BH}}=2.35\times10^7 M_{\odot}$, Bentz \& Katz 2015), assuming the same inner flow geometry for Seyfert galaxies (as inferred in De Marco et al. 2013), an X-ray reverberation soft lag would be expected at $\sim 1-2\times 10^{-4}$ Hz, and the low statistics at those frequencies do not allow us to draw strong conclusions on its presence. On the other hand, a delayed response of the warm absorber detected in these observations (Kaspi et al. 2002; Netzer et al. 2003; Krongold et al. 2003; Mao19) would be observed at lower frequencies (Silva et al. 2016), thus being either swamped by the continuum hard lags or not detectable within the analysed frequency window. 

During observations O1-O2 the lag-frequency spectrum changes notably, with the low frequency hard lags seen during U1-U3 disappearing and some tentative evidence of a soft lag (of $\sim -760\pm 660$ s) becoming apparent. The insets in Fig. \ref{fig:lagfreq} show the significant drop of intrinsic coherence (from $0.85\pm0.06$ during U1-U3 to $0.44\pm0.15$) occurring at the frequencies where the switch of lag sign between the two epochs is observed.

Although the evidence for this soft lag is marginal, its appearance in conjunction with the obscuration event might point to an association with the delayed response of the X-ray obscurer. In this event, the lag would indicate an upper limit of $\tau_{eq}\lsim 1.4$ ks, where $\tau_{eq}$ is the time needed for the gas to reach a photo-ionisation equilibrium. We notice that this value might underestimate the intrinsic response time of the gas because of the lag being diluted by the residual fraction of direct continuum flux (i.e. not modulated by the absorption features produced by the obscurer) in the soft band. However, since the obscurer absorbs a large fraction of continuum photons,
dilution has minimal impact on the lag. Indeed, following the formalism described in Mizumoto et al. (2018) and modified for the case of a partially absorbed primary continuum, it turns out that the higher the fraction of absorbed continuum, the lower the dilution of the intrinsic lag.
We estimated this fraction as $1-F^{obsc}/F^{unobsc}$ (the indices refer to the flux of the variable continuum with and without the effects of the obscuring gas) in the $0.5-2$ keV. According to our best-fit spectral model for observations O1-O2, this is of the order of $0.6-0.7$ (Table \ref{table2}). Therefore, the upper limit on the response time of the gas might be larger by a factor of $\sim 1.4-1.7$ than observed.

\section{Discussion}

Transient X-ray obscurers -- outflowing, low-ionisation gas partially eclipsing the X-ray source -- have been identified in a number of Seyfert galaxies (Risaliti et al. 2007, 2011; Nardini \& Risaliti 2011; Kaastra et al. 2014; Longinotti et al. 2013, 2019; Ebrero et al. 2016; Turner et al. 2018). In this paper we made use of X-ray spectral-timing techniques (e.g. Uttley et al. 2014) to study the fast variability of the X-ray obscurer detected in NGC 3783 in December 2016 (M17), down to the shortest possible timescales. \\
We detected variability on the part of the obscurer on timescales ranging from ten hours down to about one hour (Sect. \ref{sec:var} and Fig. \ref{fig:fvar}, right panel). We found that this variability is incoherent with the variations of the primary X-ray continuum (Sect. \ref{sec:freqresolved}), which is as expected if the properties of the absorbing gas change on these timescales (e.g. Rybicki \& Lightman 1991). We showed that variations of the ionisation state of the gas can explain the observed drop of coherence in the soft X-ray band, the most affected by absorption. In particular, in order to recover this drop of coherence, the gas should reach photo-ionisation equilibrium on timescales $<$1.5 ks, with its ionisation parameter varying within the range ${\rm{log}}\ \xi = 1.53-1.82$  on these timescales (Sect. \ref{sec:coherE} and Fig. \ref{fig:coherence} right panel). We found tentative evidence for a time delay of $\sim -760\pm 660$ s between the short timescale variations of the primary X-ray continuum and the response of the absorbing gas (Sect. \ref{sec:lagfreq} and Fig. \ref{fig:lagfreq}, right panel). This would correspond to an upper limit of $\sim 1.4$ ks on the time needed for the obscuring gas to reach a photo-ionisation equilibrium after a variation of the ionising continuum. Although the detection of the soft lag is not significant, the inferred upper limit is consistent with results obtained from our modelisation of the coherence spectrum.

\subsection{The X-ray obscurer in NGC 3783}
\label{discussion:ngc3783}

The results presented in this paper show that the X-ray obscurer in NGC 3783 varies on short timescales ranging between about one hour and ten hours. Absorption is described by a non-linear term in the radiative transfer equation (e.g. Rybicki \& Lightman 1991), whereby the intensity of an incident ray decreases exponentially, with the exponential depending on the physical properties of the gas. Therefore, absorption features are expected to vary incoherently with the ionising continuum if the physical properties of the gas change. This variability might be associated with variations of the ionisation state of the gas in response to variations of the primary continuum and/or variations of its column density (e.g. due to motions of a spatially inhomogeneous gas across our line of sight, so that the incident flux intercepts different zones of this distribution as a function of time). The results of our analysis of the $\rm{F}_{var}$, \emph{rms}, covariance, and coherence spectra of NGC 3783 during obscured epochs agree with these expectations. Indeed, the variability of the obscurer can explain: the observed distribution of fractional variability over energy (Fig. \ref{fig:fvar}, right); the observed drop of coherence in the soft band (Fig. \ref{fig:coherence}, right); the observed excess of soft band variability in the \emph{rms} spectrum with respect to the covariance spectrum (Fig. \ref{fig:covarrms}, right panels); and the marginal detection of a soft band delay (Fig. \ref{fig:lagfreq}). 
In particular, we showed that variations of the ionisation state of the gas in response to short timescale variations of the ionising continuum provide a satisfactory explanation of all the observed X-ray spectral-timing properties of the source.
Indeed, our photo-ionisation model of a X-ray obscurer can reproduce the observed drop of coherence at soft X-ray energies, provided the gas responds to the observed variations of the ionising flux in NGC 3783 on timescales $<1.5$ ks (Sect. \ref{sec:coherE}). Under this assumption, the simulated coherence spectrum closely resembles the observed drop of coherence at soft X-ray energies (Fig. \ref{fig:coherence} right panel). This decrease of coherence is ascribed to non-linear modulations of the variability of the primary continuum by variable absorption features in the soft band. The opacity of the obscuring gas is higher in the soft band, so that the drop of coherence is stronger at low energies and decreases at high energies. According to M17, the average fraction of transmitted primary continuum is $\sim$ 57 percent at $E\lsim 2$ keV, and $\sim$80 percent at $E\sim 3-4$ keV. This suggests the latter energy band to be dominated by the coherent variations of the primary continuum, thus explaining the lack of incoherence in this band.
Moreover, since the coherence is not affected by the presence of the constant scattered emission component, contrary to $\rm{F}_{var}$ spectra the effects of variable absorption can be seen down to soft energies.
The inferred short response time is also in agreement with the detected short timescale variability of the X-ray obscurer (Sect. \ref{sec:modfvar}), which implies that the obscurer in NGC 3783 should reach photo-ionisation equilibrium on timescales shorter than one hour. This inference is also supported by our tentative detection, during obscured epochs, of a delayed response in the soft band -- the most affected by absorption -- with respect to the hard X-ray photons (Fig. \ref{fig:lagfreq}, right panel). This feature was not observed during unobscured epochs, where a hard lag was instead detected (Fig. \ref{fig:lagfreq}, left panel). The concurrent disappearance of the low-frequency hard lag and the possible appearance of a low-frequency soft lag during obscured epochs hints at the variable X-ray obscurer producing a delayed response.

Given that the time variations in the soft X-ray band are dominated by the ionisation and recombination equilibration timescales in the obscurer, we can use the upper limit of $< 1.5$ ks on the recombination time of the obscurer (as inferred from our simulations, Sect. \ref{sec:coherE}) to constrain the density of the obscuring gas. To this aim, we need to estimate the recombination times of each ion, which in a highly ionised gas, generally, are given by $\tau_{rec} \sim (\alpha_{rec} n_e)^{-1} (n_i / n_{i+1})$ (Krolik \& Kriss 1995; Arav et al. 2012). Since our spectral-timing analysis does not allow us to trace timescales for individual ions, we used a photo-ionisation model of the obscurer (assuming the best-fit parameters $\rm{log}\ \xi = 1.84$, $\rm{log}\ n_H = 9.0$ and log $N_{\rm{H}} = 23.04$ as inferred in M17, K19) to compute an effective recombination timescale for the assumed gas composition, which we then rescaled to the upper limit of $< 1.5$ ks to derive a corresponding lower limit on the particle density. 
This effective recombination timescale is computed as a column-density weighted average of the recombination timescales of the different ions dominating the soft X-ray opacity, namely, the Li-like, He-like, and H-like ions of C, N, and O. For our calculations we used Cloudy and we computed ion-abundance weighted recombination rates for each of the above ions across the photo-ionised slab (using the {\tt save ionisation rates} command). The total column density is dominated by O$_{\rm{VII}}$, which comprises $\sim 40$\% of the total (with an average recombination rate of 0.014 $\rm s^{-1}$). After including all the ions in a weighted average, our model yields a mean recombination rate for the X-ray absorbing gas of 0.011 $\rm s^{-1}$, for an electron density of $n_e = 1.17 \times 10^9~\rm cm^{-3}$. Scaling this average recombination time of 91 s to our upper limit of $< 1.5$ ks sets a lower limit of $n_e > 7.1 \times 10^7~\rm cm^{-3}$ on the density of the obscurer. This agrees with the value $n_e \sim  2. 6 \times 10^{9}~\rm{cm^{-3}}$ independently inferred by M17, adopting different techniques and assumptions (and used in our simulations, Sect. 4.2). Such densities are consistent with the obscuring gas being part of the BLR.
Indeed, given the SED of observations O1 and O2 (M17), the observed X-ray flux variability (Fig. \ref{fig:lc}), and assuming the continuum varies only in normalization within each observation, changes in ionising luminosity occur within the range $L_{1-1000\ Ryd}=0.59-1.14\times 10^{44}\ \rm{erg\ s^{-1}}$ (corresponding to $\sim 0.02-0.04\ L_{\rm{Edd}}$ for a $M_{\rm{BH}}=2.35\times10^7 M_{\odot}$, Bentz \& Katz 2015). Assuming the higher density inferred by M17 and the average value of $\rm{log\ (\xi/erg\ cm\ s^{-1})} = 1.84$ (K19), this yields an upper limit of $\sim$7--10 light days on the distance of the obscuring gas from the X-ray source. Considering the size of the broad line region (BLR) in NGC 3783 at the time of the 2016 campaign (i.e. $\sim1-15$ light days, as inferred from the decomposition of the C$_{\rm{IV}}$ emission-line profile in COS-HST spectra; K19), the derived distance is consistent with the obscuring gas being inside the BLR. 
In particular, the corresponding orbital velocity at such distances is $\sim 3500-4200$ $\rm{km\ s^{-1}}$, roughly of the order of the velocity of the C$_{\rm{IV}}$ medium and broad emission components (K19). This supports the hypothesis (proposed by K19, see also Czerny \& Hryniewicz 2011) that the inner parts of the BLR, exposed to a stronger UV flux in 2016 (M17), act as a reservoir for the obscuring outflow. 
It is interesting to note that a distance of 7--10 light days (corresponding to $\sim 5200-7400\ r_g$ for a $M_{\rm{BH}}=2.35\times10^7 M_{\odot}$; Bentz \& Katz 2015), is consistent with the lower limits on the distance of the obscuring outflow detected in NGC 5548 (i.e. $\gsim 2-7$ light days, corresponding to $\gsim900-3000\ r_g$, Kaastra et al. 2014), suggesting similar launching radii.

Our simulations show that the intensity of the observed drop of coherence depends on the average ionisation state of the gas (see Sect. \ref{discussion:obsc_inSey1} and Fig. \ref{fig:sim_coherence}). Thus for a fixed value of the density, a gas located at larger (smaller) distances would be less (more) ionised and, as a consequence, the drop of coherence would result more (less) pronounced. An average value of $\rm{log\ (\xi/erg\ cm\ s^{-1})} = 1.7$ (consistent with the value of $1.84^{+0.40}_{-0.20}$ estimated by K19) accounts for the drop of coherence observed in the data of NGC 3783 (Fig. \ref{fig:coherence} right panel).

As discussed in Sect. \ref{sec:coherE}, it is difficult to reconcile the observed drop of coherence in the soft band with a partial covering model. Indeed, the fraction of continuum photons that does not intercept the obscurer would significantly increase the coherence in the soft band. This discrepancy with respect to previous modellings (e.g. M17) might be a consequence of our simulated photo-ionisation model or partial covering models oversimplifying the structure of the obscurer. In particular, our simple model assumes a single, totally covering, obscurer, while the more complex best-fit model to the time-averaged spectrum (M17) includes two partial-covering obscurers. In Sect. \ref{sec:modfvar} and Appendix \ref{sec:app2}, we show that additional variability associated with obscurer \#1, while not dominant, cannot be excluded. This variability would mostly contribute in the soft band, possibly producing the additional incoherent variability needed to counteract the increase of coherence due to continuum photons that are not intercepted by the partial-covering gas. Higher quality data and more complex modelisation would be needed to better explore this hypothesis. However, it is remarkable that despite the fact that the absorber could vary in many different ways, a very simple variation of the ionisation parameter can provide a satisfactory explanation of all the observed X-ray spectral-timing properties of the source.

\subsection{X-ray variability due to motions of the obscurer}
Variations associated with motions of the obscurer crossing the line of sight (e.g. resulting in variations of the covering factor, and possibly column density if the gas is inhomogeneous) cannot be excluded but they most likely dominate the variability on longer timescales (e.g. of the order of the duration of the obscuration event, $\sim$32 days, M17, Kaastra et al. 2018). In particular, as shown in Appendix \ref{sec:app2} and Fig. \ref{fig:fvar_cf}, strong fast variability of the covering factor can be excluded, as it does not reproduce the observed $\rm{F}_{var}$ spectrum of the source. Mild variations (by $\sim$ 2\%) cannot be excluded but they must be associated with more prominent variations of column density (Fig. \ref{fig:fvar_nhcf}) or ionisation parameter (Fig. \ref{fig:fvar}) in order to reproduce the observed peak of variability in the $\rm{F}_{var}$ spectrum. Nonetheless, given the sampled time scales (shorter than ten hours), if associated with gas motions, the observed variability would imply the gas crossing our line of sight at a significant fraction of the speed of light, that is, $\gsim 0.06c$ for a X-ray emitting region radius of $\gsim$10 $r_g$. This corresponds to a factor $\gsim 4-5$ higher than the Keplerian velocity at the estimated distance of the obscuring gas (Sect. \ref{discussion:ngc3783}), and a factor $\gsim$ 10 higher than the estimated velocity of the outflow detected in NGC 3783 (M17, K19). This conclusion is further validated after comparison of our results with the predictions reported by Gardner \& Done (2015), who studied the effects of variable occultation of the inner accretion flow due to motions of the obscuring gas on the spectral-timing properties of the source. Indeed, they show that if the gas co-rotates with the accretion flow (e.g. as expected in the case of a disk wind with a sufficiently large opening angle so as to intercept the line of sight to the X-ray source), a hard lag would be observed. Moreover, gas clumps obscuring both the soft and hard X-ray emitting parts of the accretion flow during their passage would produce coherent variability in the two bands ($\gsim$ 0.8).

\subsection{The variability properties of the X-ray continuum}
\label{discussion:continuum}

The short timescale variability properties of the hard X-ray continuum did not change significantly between the two sets of \xmm\ observations (15 years apart). Though a factor of $\sim 2$ higher level of fractional variability at $E\gsim$ 2 keV is registered during observations O1-O2 as compared to U1-U3 (see Fig. \ref{fig:fvar}), this is due to a decrease of $2-10$ keV flux mostly produced by the obscuration event. Indeed, the observed $2-10$ keV flux drops by a factor of $\sim$ 2 between the two epochs, while the intrinsic power-law flux decreases only by a factor of $\sim$ 1.2 (Table \ref{table2}). 
This slight decrease is within the expected scatter of average flux produced by red noise fluctuations associated with the underlying variability process (Vaughan et al. 2003). After accounting for this scatter, we obtain $<\rm{F}_{var}>_{2-10 keV}=7.6^{+2.5}_{-2.1}$ percent during U1-U3 and $<\rm{F}_{var}>_{2-10 keV}=14.4^{+8.2}_{-6.8}$ percent during O1-O2\footnote{The procedure to estimate the scatter in the $\rm{F}_{var}$ associated with red noise fluctuations is described in Vaughan et al. (2003) and Ponti et al. (2012). Here we made use of the constraints on the PSD derived by Markowitz et al. (2003) and Summons et al. (2007) from long-term {\it RXTE} observations.}. The two values are in broad agreement within the (90 per cent confidence level) uncertainties. Therefore, we conclude that there is no strong evidence for non-stationarity of the underlying stochastic process during the analysed periods, while weak stationarity over timescales of 15 years is a reasonable assumption for Seyfert galaxies (e.g. considering the timescales for non-stationarity in BH X-ray binaries and scaling them to Seyferts). In other words (considering also results from X-ray spectral analysis; M17), while the UV flux changes significantly between the two epochs (M17), we do not observe any major change in the properties of the X-ray continuum emission that might be related to the triggering of the obscuration event.

The lag-frequency spectrum of observations U1-U3 shows the presence of a low-frequency hard lag with a power-law dependence on frequency (e.g. Papadakis et al. 2019; Epitropakis \& Papadakis 2017), a common feature of unobscured AGN (De Marco et al. 2013; Walton et al. 2013; Kara et al. 2016). These lags are characterised by high levels of intrinsic coherence, in some cases, they are mildly decreasing with increasing difference of photon energies (Ar\'{e}valo et al. 2008; Epitropakis \& Papadakis 2017). This behaviour is in fact seen in the data of NGC 3783, during observations U1-U3 (Fig. \ref{fig:coherence}, left panel), when the line of sight to the X-ray source is not affected by obscuration.

Some of the proposed explanations for the low-frequency hard lags and the high coherence involve mass accretion rate fluctuations propagating through a radially extended and spectrally inhomogeneous X-ray source (e.g. Kotov et al. 2001; Ar\'{e}valo \& Uttley 2006), or fluctuations propagating through the disc and producing correlated variations of power-law index as a consequence of coronal X-ray heating (Uttley et al. 2014). In these scenarios, the hard lags and the high coherence are intrinsic properties of the X-ray continuum. In the context of the best-fit spectral model of M17 for NGC 3783, these spectral-timing properties can be interpreted in terms of fluctuations propagating through the soft and then through the hard Comptonisation regions.

An alternative model to explain the hard lags in AGN invokes scattering in low-ionisation, optically thick circumnuclear material located at large distances from the BH which does not intersect the line of sight to the source (Miller et al. 2010; Turner et al. 2017). However, this latter model appears disfavoured in that this distant X-ray reflector should be variable on the sampled timescales and, therefore, significantly contributing to the $\rm{F}_{var}$ and frequency-resolved spectra of the source. However, we found clear evidence for a narrow drop of variability at $E=6.4$ keV in the $\rm{F}_{var}$ spectrum (Fig. \ref{fig:fvar}) which can be fully explained by neutral Fe K$\alpha$ emission, constant on timescales shorter than ten hours. These timescales imply that the distance of the reflecting gas responsible for this feature is $\gsim 10^{15}\ \rm{cm}$, which is consistent with being produced at the same distance of the BLR or beyond (K19).

\begin{figure*}
        \includegraphics[angle=0,width=0.95\columnwidth]{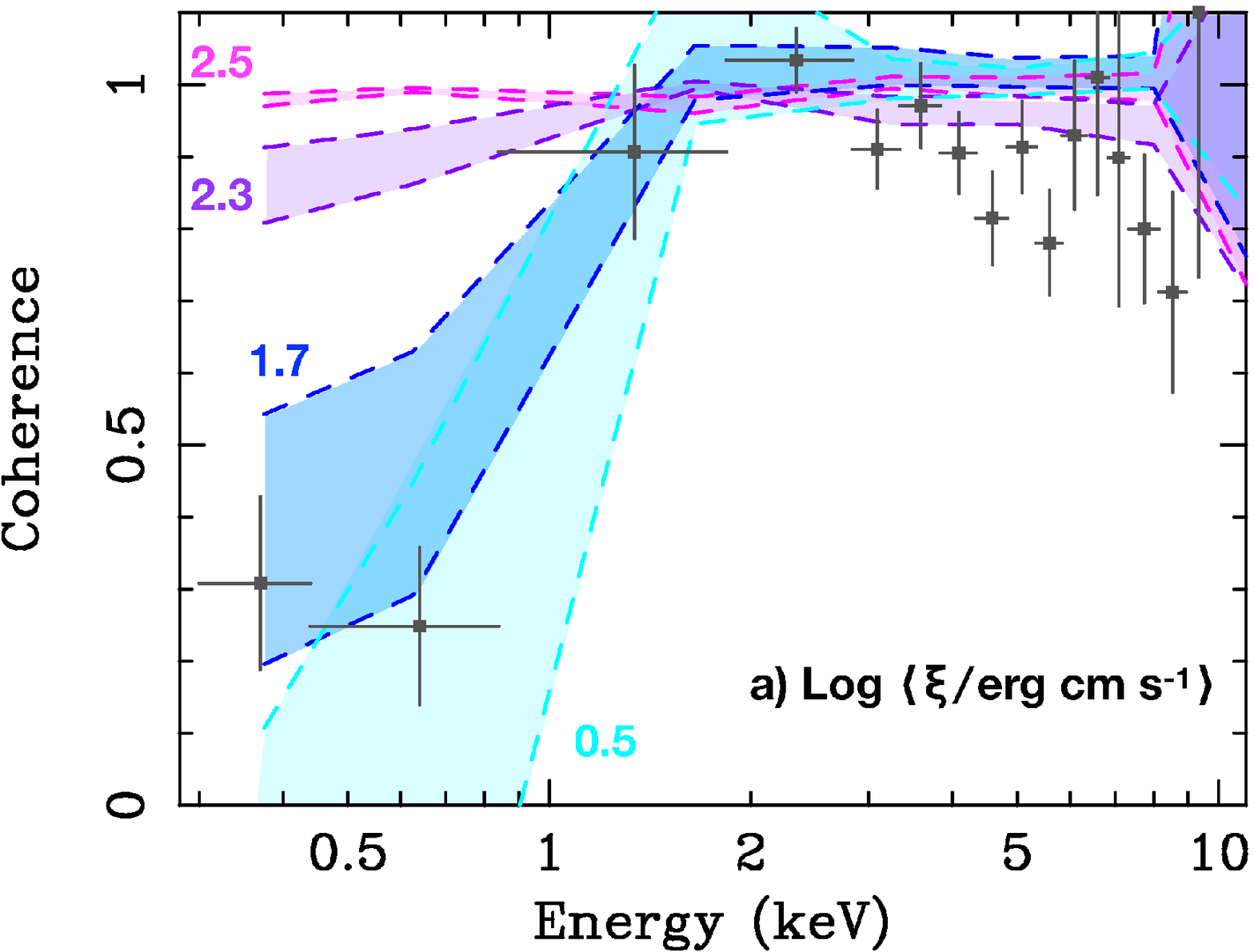} 
        \hspace{0.6cm}
        \includegraphics[angle=0,width=0.95\columnwidth]{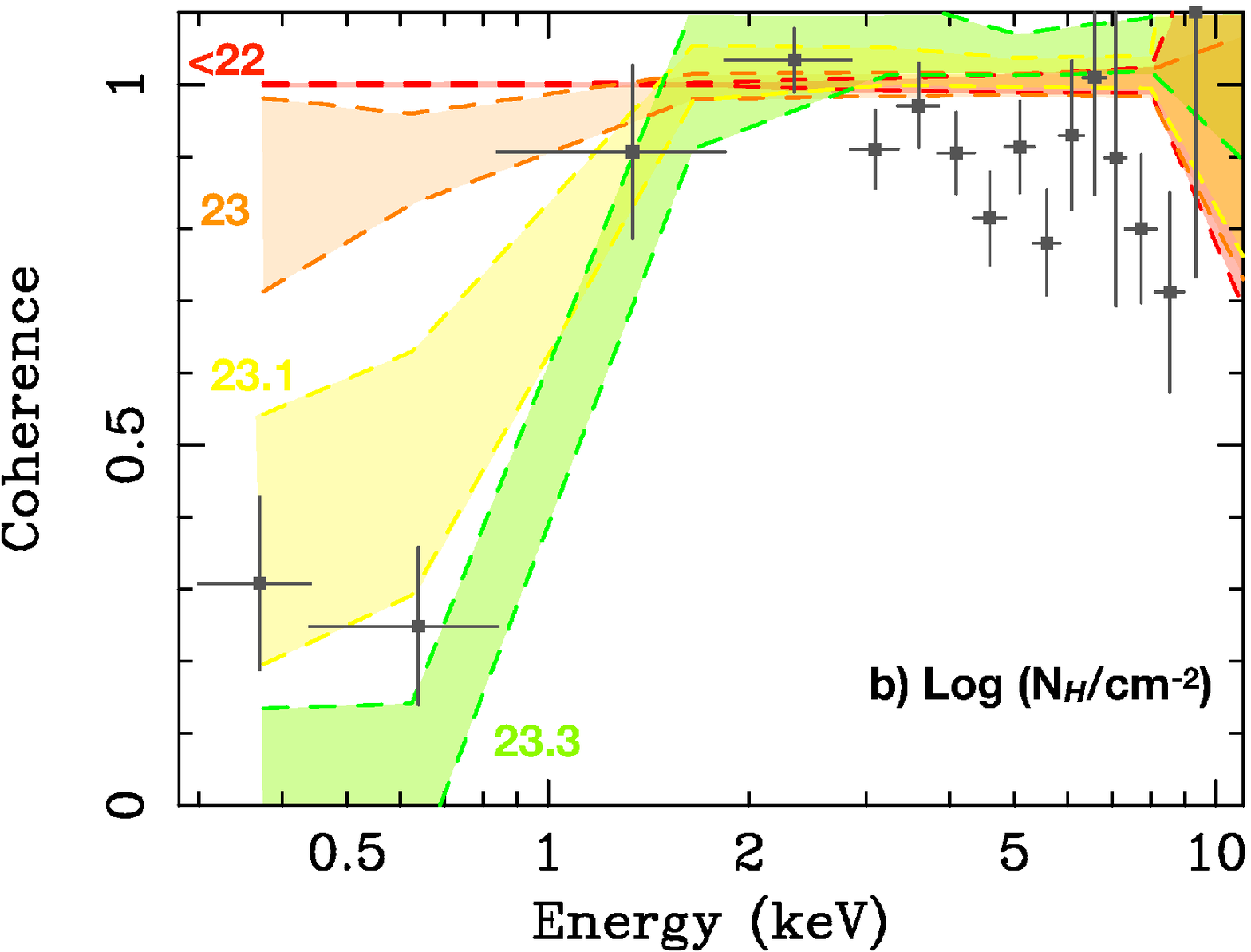}
    \caption{Simulated coherence spectra for an obscuring gas responding to variations of the ionising continuum. Panel a) dependency of the intensity of the drop of coherence on the average ionisation parameter: the colours indicate different degrees of ionisation of the gas, with ${\rm{log}}\ \xi$ varying within the ranges ${\rm{log}}\ \xi=0.33-0.61$ (cyan), ${\rm{log}}\ \xi=1.53-1.82$ (blue), ${\rm{log}}\ \xi=2.14-2.42$ (purple), ${\rm{log}}\ \xi=2.33-2.61$ (magenta). These respectively correspond to distances of $d=$ 40, 10, 5, and 4 light days of the gas from the X-ray source. Panel b) dependency of the intensity of the drop of coherence caused by variations of the ionisation parameter (within the range ${\rm{log}}\ \xi=1.53-1.82$), on the average column density of the obscurer: the different values of column density are obtained by changing the depth of the gas, $\Delta d =3.8\times10^{12} \rm{cm}$ (red), $3.8\times10^{13} \rm{cm}$ (orange), $4.8\times10^{13} \rm{cm}$ (yellow), and $7.7\times10^{13} \rm{cm}$ (green). For comparison with the observations, the measured coherence during O1-O2 is overplotted (grey squares) on the simulated coherence spectra.}
    \label{fig:sim_coherence}
\end{figure*}

\subsection{The effects of variable obscuration on the X-ray spectral-timing properties of active galactic nuclei}
\label{discussion:obsc_inSey1}

The results presented in this paper can be generalised for a discussion on the effects of variable obscurers on the X-ray spectral-timing properties of AGN. The observed variability timescales can be used to put constraints on the density of the gas and thus on its distance (Nicastro et al. 1999; Krongold et al. 2007). However, the denser the gas, the shorter the variability timescales involved, possibly as short as those commonly associated with emission processes from the inner accretion flow. The use of spectral-timing techniques enables studying obscurers variability on very short timescales; thus, providing constraints on the density of gas components located closer to the BH. 
Given the similar timescales involved, identifying the spectral-timing signatures of variable obscurers is critical in order to correctly disentangle them from those intrinsic to the X-ray source.\\

\noindent{\bf X-ray coherence:} One of the main expected signatures of variable absorption is a decrease of coherence as a consequence of the transfer equation for absorption being non-linear (see dicussion in Sect. \ref{discussion:ngc3783}). Silva et al. (2016) built time-dependent photo-ionisation models (Nicastro et al. 1999; Kaastra et al. 2012) to study the X-ray spectral-timing signatures of variable warm absorbers responding to changes of the ionising continuum and applied them to the case of NGC 4051. They reported high levels of coherence associated with warm absorbers responding to the variability of the ionising continuum. However, this result should depend on the properties of the absorbing gas, and, in particular, on the amount of absorbed flux. Therefore, we ran simulations to test a wider range of parameter space for the physical properties of the gas, so as to encompass the case of X-ray obscurers. Following the procedure described in Sect. \ref{sec:coherE}, we used Cloudy to obtain synthetic coherence spectra for changes of the ionisation state of the gas in response to primary continuum variations and assuming different properties of the gas. 
For the purposes of simplicity, we assumed that the response time of the obscuring gas is shorter than the timescales sampled. In other words, the integration time of each simulated spectrum is longer than the response time of the gas, so that each spectrum corresponds to an equilibrium configuration. 
We assumed the same particle density of the obscurer in NGC 3783 ($n_e=2.6\times 10^9 \rm{cm^{-3}}$) and the same ranges of ionising continuum luminosities (Sect. \ref{sec:coherE}). We tested the two following cases. 

Case a) \emph{dependency of the intensity of the drop of coherence on the average ionisation parameter of the absorbing gas:} the particle density ($n_e$) and depth of the absorbing gas ($\Delta d$) are fixed, so that log $N_{\rm{H}} = 23.1$, while different distances in the range $d=4-40$ light days from the ionising source are tested. These distances roughly correspond to the estimated distance of the BLR in sources of $M_{BH}\sim 10^7 \rm{M_{\odot}}$ (e.g. Peterson 2006). Given that $\xi=L/(n_ed^2)=L\Delta d/(N_{\rm{H}}d^2)$, for each choice of the distance there is a one-to-one relation between $\xi$ and $L$. Indeed, in these simulations, given the observed variations of luminosity $L$, the ionisation parameter varies within the ranges ${\rm{log}}\ \xi=2.33-2.61$ for d=4 light days, and ${\rm{log}}\ \xi=0.33-0.61$ for d=40 light days.

Case b) \emph{dependency of the intensity of the drop of coherence on the column density of the absorbing gas}: the absorbing gas has a fixed distance ($d=10$ light days) and density (thus, given the observed variations of luminosity, the ionisation parameter varies within the range ${\rm{log}}\ \xi=1.53-1.82$), and column densities in the range $N_{\rm{H}}=10^{22-23.3}\rm{cm^{-2}}$ as a consequence of considering different depths of the absorbing gas (between $\sim 0.4-7.7\times10^{13}\ \rm{cm}$).

Results are shown in Fig. \ref{fig:sim_coherence}. As is clear from these simulations, and in line with theoretical expectations, the average ionisation parameter and the depth of the gas determine the intensity of the drop of coherence.
The latter appears at the energies where the gas is more opaque to the ionising radiation (particularly in the soft band), absorbing, thus, a larger fraction of continuum flux. This is in line with the results from Silva et al. (2016), who also reported a slight decrease of coherence in more absorbed bands.
Since the gas opacity in a given energy band depends on the ionisation state, lower values of the ionisation parameter cause a more intense drop in the soft band. These simulations show that variable X-ray obscurers (characterised by typically high column densities and low ionisation parameter) would produce significant effects on the observed coherence, provided the analysed timescales are longer than the response time of the gas. Warm absorbers may decrease the coherence, although the strength of their contribution depends on their specific parameters. In the case of NGC 3783, the column densities of the dominant warm absorber components are in the range of $N_{\rm{H}}\sim 10^{21-22}~\rm{cm^{-2}}$ and their ionisation parameter in the range ${\rm{log}}\ \rm{(\xi/ erg\ cm\ s^{-1})}\sim2-3$ (Mao19). As can be inferred from Fig. \ref{fig:sim_coherence}, these components are not expected to produce a significant decrease of coherence.

\noindent{\bf X-ray time lags:} X-ray time lags provide an additional important diagnostic of variability associated with absorbing gas. 
Silva et al. (2016) showed that if absorption variability is due to changes of the ionisation state of the gas, then the most absorbed (soft) X-ray bands would show a relative delay with respect to the least absorbed (hard) X-ray bands. The resulting (soft) lag depends on the response time of the gas, thus on its electron density (as $\tau_{eq}\propto n_{e}^{-1}$, Peterson 2006), which ultimately determines the distance from the X-ray source (as $d \propto n_{e}^{-1/2}$) for a given ionisation parameter. As discussed in Silva et al. (2016), one of the main problems in extracting the intrinsic response time of the gas from the measured lag is related to dilution effects, due the presence of direct continuum emission in the same energy band of the absorber. 
As an example, assuming the parameters of the warm absorber detected in NGC 4051, Silva et al. (2016) predicted a maximum soft lag of 100 sec, much shorter than the maximum delay estimated for the obscurer in NGC 3783 (Sect. \ref{sec:lagfreq}), despite the former being associated with a less dense gas. This is due to the fact that a lower fraction of soft band flux is modulated by changes of the ionisation state of the warm absorber in NGC 4051. Indeed, this gas has lower $N_{\rm{H}}$ and is more ionised than the X-ray obscuring gas in NGC 3783, thus it takes away a lower fraction of primary X-ray flux as compared to the obscurer in NGC 3783.
As a consequence, the resulting soft lag is significantly more diluted. In the case of NGC 3783, the high column density and low ionisation parameters of the obscurer ensure that absorption variability modulates a larger fraction of soft band variability, so that the expected dilution is small (Sect. \ref{sec:lagfreq}).\\

\begin{figure}
        \includegraphics[angle=0,width=0.95\columnwidth]{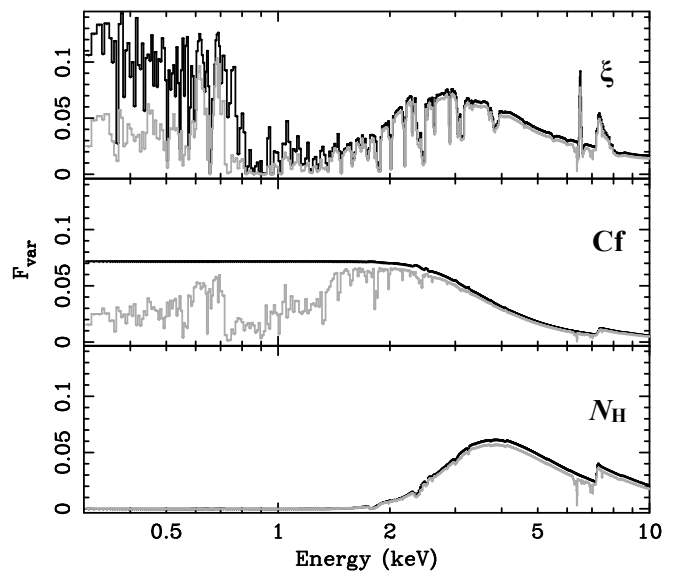} 
    \caption{Simulations of high resolution $\rm{F}_{var}$ spectra, assuming variations of single parameters (${\rm{log}}\ \xi$ varying by $\sim$ 24\%,  covering factor varying by $\sim$4\%, and $N_{\rm{H}}$ varying by $\sim$ 19\%-23\%, as indicated in the upper corner of each plot) of obscurer \#2 in NGC 3783. The black and light grey curves are, respectively, the $\rm{F_{var}}$ spectra excluding and including the contribution from the constant scattered emission and reflection components.}
    \label{fig:fvar_highres}
\end{figure}

\noindent{\bf Fractional variability:} 
The effects of absorption variability on the $\rm{F}_{var}$ spectrum have been extensively discussed in the literature (e.g. Mizumoto \& Ebisawa 2017; Parker et al. 2017b, 2020). As illustrated in Fig. \ref{fig:fvar_highres} through simulations of $\rm{F}_{var}$ spectra obtained by letting single parameters of the obscurer vary (following the procedure described in Sect. \ref{sec:modfvar}), variations of the depth of the most prominent absorption lines manifest as discrete peaks in the $\rm{F}_{var}$ spectrum. These, however, would appear strongly smoothened in low resolution $\rm{F}_{var}$ spectra, as seen in Fig. \ref{fig:fvar}.  A smoother spectral shape is expected in the case of variability of the column density and of the covering factor of low ionisation and high column density obscuring gas (Fig. \ref{fig:fvar_highres}, middle and lower panels). 
Additional contribution from constant emission components would reduce the fractional variability, producing dips (at the energy of constant emission lines), or broad variability drops (either due to constant continuum emission or to unresolved emission lines). 
The effect of these constant components on high resolution $\rm{F}_{var}$ spectra is shown in Fig. \ref{fig:fvar_highres} (light grey curves) for the case of NGC 3783. 
A combined analysis of high resolution time-averaged and $\rm{F}_{var}$ spectra would enable the identification of the main parameters causing variability of the obscuring gas. This will be attainable with the future missions XRISM and Athena (e.g. Guainazzi \& Tashiro 2018).

\section{Conclusions}
Given their low ionisation parameter, X-ray obscurers take away a significant fraction of primary X-ray flux. If the physical properties of the obscurer vary on short timescales, then a significant decrease of coherence is expected. This is confirmed by observations of the X-ray obscurer in NGC 3783, which is found to vary incoherently on timescales ranging between about one hour to ten hours. This variability is likely associated with changes of the ionisation state of the gas in response to variations of the ionising continuum. The gas responds on timescales $<$1.5 ks, which corresponds to a lower limit on the particle density of the gas of $n_e> 7.1\times10^{7} {\rm{cm^{-3}}}$. Such densities are consistent with the obscuring gas being located within the BLR.

\begin{acknowledgements}
This work is based on observations obtained with \xmm\, an ESA science mission with instruments and contributions directly funded by ESA Member States and NASA. This research has made use of data obtained with the \nustar\ mission, a project led by the California Institute of Technology (Caltech), managed by the Jet Propulsion Laboratory (JPL) and funded by NASA. BDM acknowledges support from the European Union's Horizon 2020 research and innovation programme under the Marie Sk{\l}odowska-Curie grant agreement No. 798726, and grant agreement No. 665778 via the Polish National Science Center grant Polonez 2016/21/P/ST9/04025. TPA acknowledges support from Polish National Science Center grant Polonez 2016/21/P/ST9/04025 and Preludium 2016/21/N/ST9/0331. SB acknowledges financial support from ASI under grants ASI-INAF I/037/12/0 and n. 2017-14-H.O. AGM acknowledges partial support from
 Nardowe Centrum Nauki (NCN) award 2016/23/B/ST9/03123. JM is supported by STFC (UK) through the University of Strathclyde UK APAP network grant ST/R000743/1. DJW acknowledges support from STFC in the form of an Ernest Rutherford fellowship. CP is supported by European Space Agency (ESA) Research Fellowship. POP thanks financial support from the CNES french agency and the CNRS/PNHE. The authors thank the referee for their helpful comments.
\end{acknowledgements}

%
%

 \begin{appendix} \section{best fit time-averaged spectral model}
\label{sec:app}
The baseline model used for the simulations of the $\rm{F}_{var}$ spectra in Sect. \ref{sec:modfvar} is the best-fit model derived in M17 and Mao19 from the analysis of \xmm\ , \nustar\ , and \chandra\  data, obtained using the SPEX package (Kaastra et al. 1996). We translated this model into an Xspec model (computing Cloudy tables for absorption and scattered emission components) and applied it to the EPIC pn data. The most relevant parameters inferred from these fits are summarised in Table \ref{table2}. 

\begin{table}
        \centering
        \caption{Baseline model used for simulations of $\rm{F}_{var}$ spectra.}
         \label{table2}
         \begin{tabular}{lccc} 
         \hline
          \noalign{\smallskip}
                                                                    &         U1-U3         &     O1    &    O2      \\
         \noalign{\smallskip}                                                       
          \hline
          \hline
           \noalign{\smallskip}

        ${\rm{log}}\ N_{\rm{H,Gal}}$                                 &    \multicolumn{3}{c}{$20.98^{(f)}$}   \\
         \noalign{\smallskip}
        \hline
        \noalign{\smallskip}
                                                                      \multicolumn{4}{c}{SCATTERED EMISSION}  \\   
${\rm{log}}\ N_{\rm{H1}}$                      &            $23.78^{(f)}$                &   \multicolumn{2}{c}{$23.4^{(f)}$}\\
        
        ${\rm{log}}\ \xi_{1}$                               &      $2.60^{(f)}$              &     \multicolumn{2}{c}{$2.58^{(f)}$}   \\
        
${\rm{log}}\ N_{\rm{H2}}$                      &             $22.72^{(f)}$                 &   \multicolumn{2}{c}{$22.48^{(f)}$} \\
        
        ${\rm{log}}\ \xi_{2}$                               &      $1.35^{(f)}$              &     \multicolumn{2}{c}{$1.03^{(f)}$}  \\
        
         \noalign{\smallskip}
        \hline
        \noalign{\smallskip}
                                                                      \multicolumn{4}{c}{WARM ABSORBER}  \\   
${\rm{log}}\ N_{\rm{H1}}$                      &     $22.40\pm0.04$        &  \multicolumn{2}{c}{$21.92^{+0.31}_{-0.11}$}  \\
        
        ${\rm{log}}\ \xi_{1}$                      &      $3.08\pm0.01$        &   \multicolumn{2}{c}{$2.60^{+0.11}_{-0.13}$}     \\
        
${\rm{log}}\ N_{\rm{H2}}$                      &     $21.74\pm0.01$  &    \multicolumn{2}{c}{$22.33\pm0.04$}  \\
        
        ${\rm{log}}\ \xi_{2}$                     &   $1.06_{-0.04}^{+0.01}$    &  \multicolumn{2}{c}{$1.67^{+0.06}_{-0.15}$}          \\
        
${\rm{log}}\ N_{\rm{H3}}$                      &  $22.32\pm0.02$    &    \multicolumn{2}{c}{$22.50^{+0.06}_{-0.02}$} \\
        
        ${\rm{log}}\ \xi_{3}$                      &  $2.56\pm0.02$     &        \multicolumn{2}{c}{$2.34^{+0.02}_{-0.05}$}   \\
        
         \noalign{\smallskip}
        \hline
        \noalign{\smallskip}
                                                                      \multicolumn{4}{c}{CUTOFFPL}  \\                                                                                          
        $\Gamma$                                            &     $1.75\pm 0.01$ &   $1.90\pm 0.02$ & $1.94\pm0.03$   \\   
        $E_{cut}$                                                    & \multicolumn{3}{c}{340} \\
        \noalign{\smallskip}
                                                                       \multicolumn{4}{c}{0.3-2 keV}  \\        
        $F^{obsc}$ &        -                    &   1.1                   &  2.2                   \\
        $F^{unobsc}$ &      4.1                  &   4.1                   &  5.3                   \\
        \noalign{\smallskip}    
                                                                               \multicolumn{4}{c}{2-10 keV}  \\  
         $F^{obsc}$ &           -                    &   2.6                  &  3.4                   \\      
         $F^{unobsc}$ &      5.3                  &   4.1                   &  4.9                   \\
          \noalign{\smallskip}
        \hline
         \noalign{\smallskip}
                                                                      \multicolumn{4}{c}{COMPTT}  \\
         $T_{seed}$            &       $1.47^{(f)}$      &    \multicolumn{2}{c}{$1.08^{(f)}$}          \\
         $T_c$                     &       $0.54^{(f)}$     &    \multicolumn{2}{c}{$0.13^{(f)}$}          \\
         $\tau_{depth}$        &       $9.9^{(f)}$      &     \multicolumn{2}{c}{$22^{(f)}$}       \\
         \noalign{\smallskip}
         
                                                                       \multicolumn{4}{c}{0.3-2 keV}  \\        
        $F^{obsc}$ &          -                 &   0.3                     & 0.8                       \\
         \noalign{\smallskip}   
                                                                     \multicolumn{4}{c}{2-10 keV}  \\        
        $F^{unobsc}$ &        1.5                &  1.0                     & 2.0                       \\
         \noalign{\smallskip}
                      \hline
                       \noalign{\smallskip}
                                                                      \multicolumn{4}{c}{OBSCURER \#1}  \\
        $Cf_{1}$                                                &             -                 &  $ 0.47^{(f)}$  &  $ 0.38^{(f)} $  \\

        ${\rm{log}}\ N_{\rm{H1}}$                      &              -                 &  $22.92\pm0.04$ & $22.17\pm0.13$ \\

        ${\rm{log}}\ \xi_{1}$                                &             -                  & $1.84^{(f)} $           & $1.84^{(f)} $          \\
         \noalign{\smallskip}
 \hline  
  \noalign{\smallskip}     
                                                                      \multicolumn{4}{c}{OBSCURER \#2}  \\        
        $Cf_{2}$                                                &              -                 &   $0.51 (f)$             & $0.48^{(f)}$        \\

        ${\rm{log}}\ N_{\rm{H2}}$                      &              -                  &   $23.32\pm0.03$ & $23.40\pm0.04$ \\
        
        ${\rm{log}}\ \xi_{2}$                               &               -                 &   $1.84^{(f)} $          &  $1.84^{(f)} $  \\
        
  \noalign{\smallskip}       
        \hline
         \noalign{\smallskip}
        $\chi^2/d.o.f.$                                       & $2676/1883$  &      \multicolumn{2}{c}{$3965/3662$ }  \\
         \noalign{\smallskip}
                \hline
                 \noalign{\smallskip}
        \end{tabular}
\tablefoot{Fluxes are in units of $\rm{10^{-11}\ erg s^{-1} cm^{-2}}$ ($F^{obsc}$ and $F^{unobsc}$ refer, respectively, to fluxes with and without the effects of the obscurer), power-law cut-off ($E_{cut}$) in units of keV, seed photon temperature ($T_{seed}$) in units of eV, plasma temperature ($T_c$) in units of keV, column densities ($N_{\rm{H}}$) in units of $\rm{cm^{-2}}$, ionisation parameters (defined as $\xi=L_{1-1000\ Ryd}/n_{e}r^{2}$, where $n_{e}$ is the electron density, $r$ the distance of the absorbing gas, and $L_{1-1000\ Ryd}$ the ionising luminosity) in units of $\rm{erg\ cm\ s^{-1}}$, \emph{(f)} fixed to values obtained by Mao19 and M17.}
\end{table}

\section{tests of $\rm{F}_{var}$ models}
\label{sec:app2}

\begin{figure}
        \includegraphics[angle=0,width=0.95\columnwidth]{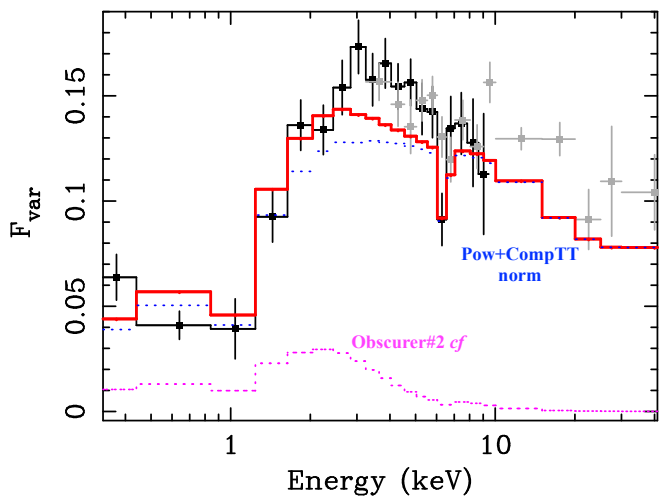} 
    \caption{Variations of the covering factor of obscurer \#2 on top of variations of the primary continuum.} 
    \label{fig:fvar_cf}
\end{figure}

\begin{figure}
        \includegraphics[angle=0,width=0.95\columnwidth]{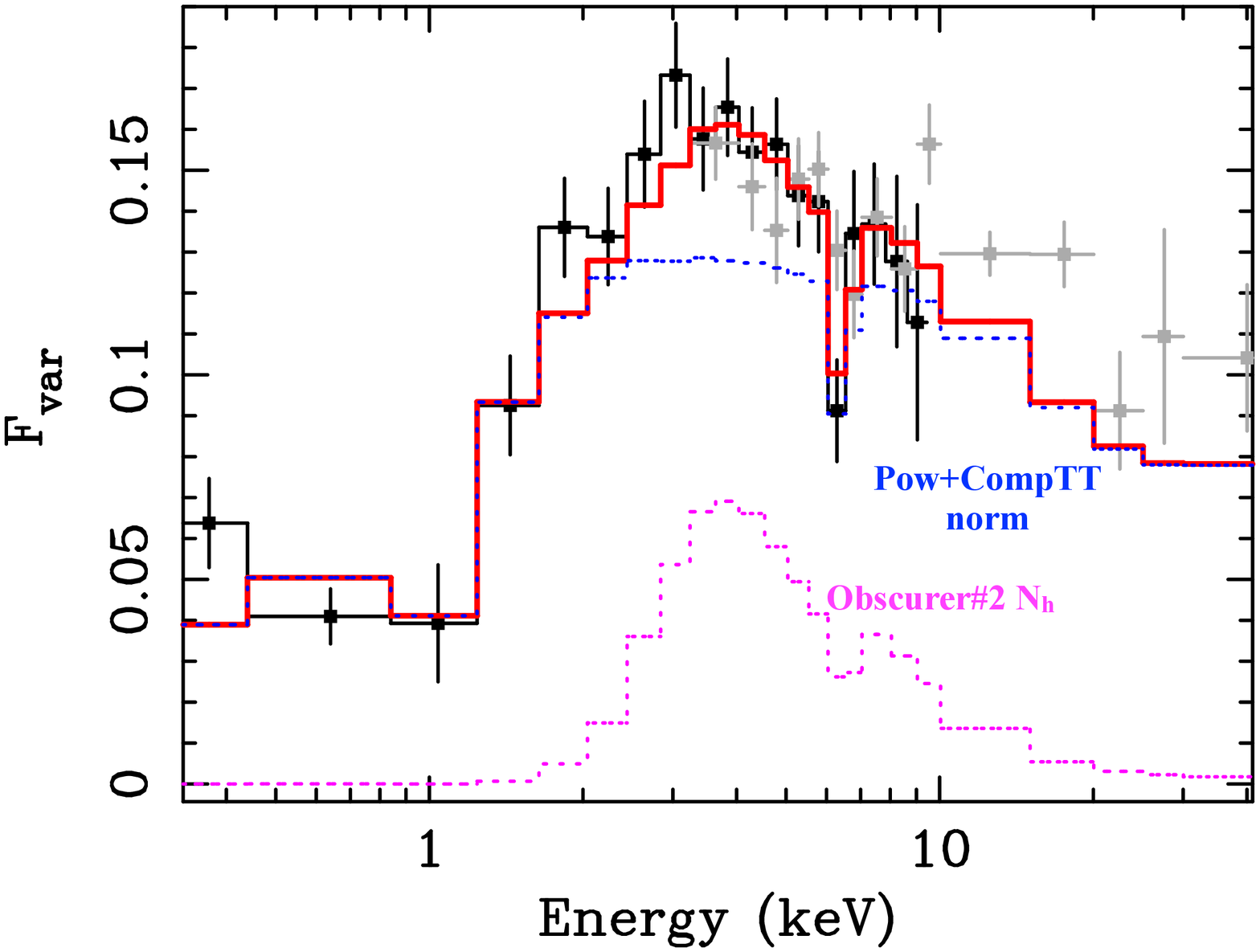} 
    \caption{Variations of the column density of obscurer \#2 on top of variations of the primary continuum.} 
    \label{fig:fvar_nh}
\end{figure}

\begin{figure}
        \includegraphics[angle=0,width=0.95\columnwidth]{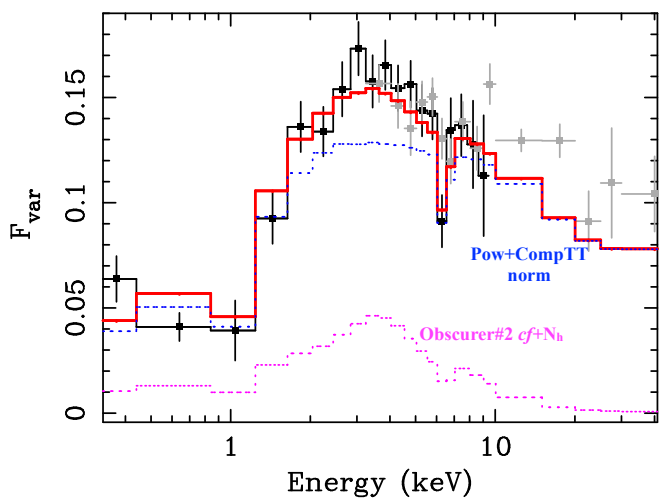} 
    \caption{Simultaneous variations of column density and covering factor of obscurer \#2 on top of variations of the primary continuum.} 
    \label{fig:fvar_nhcf}
\end{figure}

\begin{figure}
        \includegraphics[angle=0,width=0.95\columnwidth]{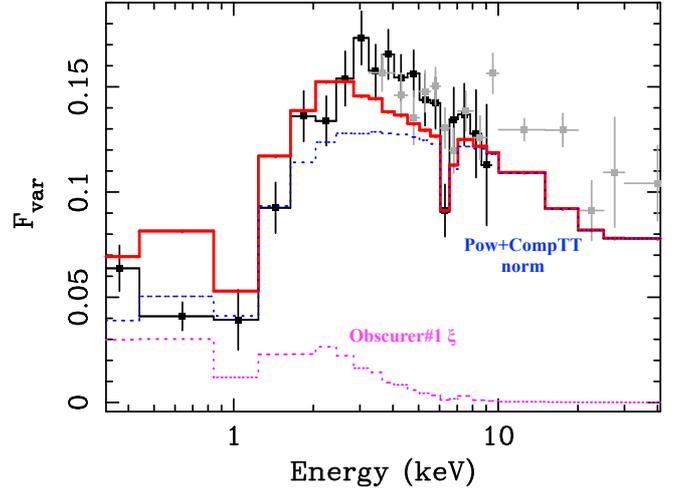} 
    \caption{Variations of the ionisation parameter of obscurer \#1 on top of variations of the primary continuum.} 
    \label{fig:fvar_O1xi}
\end{figure}

\begin{figure}
        \includegraphics[angle=0,width=0.95\columnwidth]{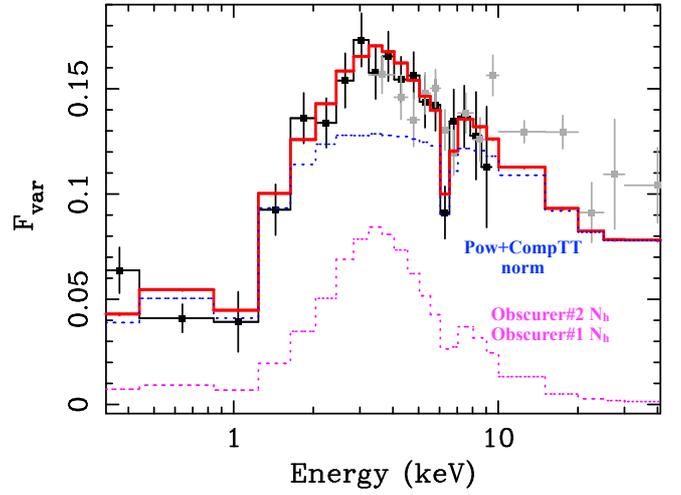} 
    \caption{Simultaneous variations of the column density of obscurer \#1 and \#2 on top of variations of the primary continuum.} 
    \label{fig:fvar_O2nhO1nh}
\end{figure}

This section discusses the additional models for the observed $\rm{F}_{var}$ spectrum of O1-O2 that have been tested. Variations of single or a combination of parameters of the two obscurers were considered on top of the variations of the primary continuum considered in Sect. \ref{sec:modfvar}.\\
\emph{Variable covering factor of obscurer \#2}: variations of the covering factor of obscurer \#2 alone cannot explain the observed spectral shape of the $\rm{F}_{var}$ in that such variations would produce a peak of variability at energies lower than observed. Fig. \ref{fig:fvar_cf} shows simulations obtained for variations of the covering factor of 2\%. Increasing the variability amplitude in an attempt to model residuals at $E\sim 2-8$ keV results in overpredicting the variability amplitude at $E\lsim 2$ keV.\\
\emph{Variations of column density of obscurer \#2}: variations of the column density of obscurer \#2 (of the order of 22--29\%) can reproduce the observed spectral shape, although this model predicts a slight mismatch between the observed and predicted peak of variability (Fig. \ref{fig:fvar_nh}). This can be compensated by mild variations of the covering factor of obscurer \#2 or of the parameters of obscurer \#1 (see below), as both produce peak variability at slightly softer energies.\\
\emph{Simultaneous variations of column density and covering factor of obscurer \#2}: the slight shift of peak variability observed when letting only the column density of obscurer \#2 vary can be compensated by mild (by 2\%) additional variations of the covering factor (Fig. \ref{fig:fvar_nhcf}).\\
\emph{Variability of obscurer \#1}: given its lower average column density (see Table \ref{table2}) obscurer \#1 would contribute to the observed variability mostly at energies $E\lsim 2-3$ keV, slightly lower than the peak of variability observed in the data. Fig. \ref{fig:fvar_O1xi} shows an example obtained by letting the ionisation parameter of obscurer \#1 vary as $\xi \propto F_{pow}$.\\
\emph{Simultaneous variations of column density of obscurer \#2 and parameters of obscurer \#1}: the slight shift of peak variability observed when letting only the column density of obscurer \#2 vary can also be compensated for by additional variations of obscurer \#1, as this mostly contributes to variability at lower energies (Fig. \ref{fig:fvar_O1xi}). In Fig. \ref{fig:fvar_O2nhO1nh}, we show as an example, the model obtained by letting the column density of both obscurers vary (by 18--23\% for obscurer \#1 and by 22--29\% for obscurer \#2). However, variations of the column density of obscurer \#2 combined with mild variations of the covering factor (by $\sim$ 2\%) of obscurer \#1 can   reproduce the observed data just as well.\\

\end{appendix}

\end{document}